\documentclass[prb,twocolumn,showpacs,amsmath,amssymb,floatfix]{revtex4}
\usepackage{ifpdf}
\ifpdf
  \usepackage[pdftex]{graphicx}
  \DeclareGraphicsExtensions{.pdf,.pdftex,.png,.jpg,.mps}
\else
  \usepackage[dvips]{graphicx}
  \DeclareGraphicsExtensions{.eps}
\fi

\usepackage[ngerman,english]{babel}

\usepackage{nicefrac}

\usepackage{pifont}

\newif\ifhyper
\hypertrue

\ifhyper
\usepackage{hyperref}
\hypersetup{%
  a4paper = {true},
  bookmarksnumbered = {true}, 
  bookmarksopen = {true}, 
  bookmarksopenlevel = {2}, 
  breaklinks = {true}, 
  colorlinks = {true}, 
  pdfauthor = {\textcopyright\ Miha\l Karski, Carsten Raas, and G\"otz S. Uhrig},
  pdfcreator = {\LaTeX\ with package \flqq hyperref\frqq},
  pdfkeywords = {metal insulator transition, MIT, Hubbard model, infinite
    dimension, dynamic mean-field theory, DMFT, DMRG, D-DMRG}
  pdfmenubar = {true},
  pdfstartpage = {1},
  pdfstartview = {FitH}, 
  pdfsubject = {Single-particle dynamics in the vicinity of the Mott-Hubbard metal-to-insulator transition},
  pdftitle = {Single-particle dynamics in the vicinity of the Mott-Hubbard metal-to-insulator transition},
  pdftoolbar = {true},
  plainpages = {false}, 
  urlcolor = {blue} 
}
\fi

\newif\ifadr
\adrtrue

\begin{document}

\title{Single-Particle Dynamics in the Vicinity of the Mott-Hubbard 
Metal-to-Insulator Transition}

\ifadr

\author{Micha{\l} Karski}
\email[\emph{Electronic address: }]{karski@uni-bonn.de}
\affiliation{Institut f\"ur Angewandte Physik der Universit\"at Bonn,
  Wegelerstr.~8,
  53115 Bonn,
  Germany}

\author{Carsten Raas}
\email[\emph{Electronic address: }]{carsten.raas@uni-dortmund.de}

\author{G\"otz S.~Uhrig}
\email[\emph{Electronic address: }]{goetz.uhrig@uni-dortmund.de}
\homepage[\\\emph{Homepage: }]{http://t1.physik.uni-dortmund.de/uhrig/}
\affiliation{Lehrstuhl f\"ur Theoretische Physik I,
  Otto-Hahn-Stra\ss{}e 4,
  Technische Universit\"at Dortmund,
  D-44221 Dortmund,
  Germany}

\fi

\date{\today}

\begin{abstract}
The single-particle dynamics close to a metal-to-insulator
transition induced by strong repulsive interaction between the electrons
is investigated. The system is described by a half-filled Hubbard model
which is treated by dynamic mean-field theory evaluated by high-resolution
dynamic density-matrix renormalization. We provide theoretical spectra
with momentum resolution which facilitate the comparison to photoelectron
spectroscopy.
\end{abstract}



\maketitle


\section{Introduction}
\label{sec:introduction}


Strongly correlated systems are notoriously difficult to understand.
In particular, this is true when parameter regions are considered where
the system changes its behavior qualitatively, for instance where new
excitations emerge. This is precisely the situation around the 
metal-to-insulator transition of electrons which repel each other
strongly on a tight-binding lattice due to the Coulomb interaction.

We consider a half-filled lattice with  
one electron per site on average. For strong on-site repulsion, it is 
energetically forbidden that two electrons occupy the same site.
Because there is one electron per site on average this implies that
there is precisely one electron per site. No electron can move so that the
system is insulating. The remaining degree of freedom is the orientation
of the electron spins. This freedom gives rise to collective magnetic
excitations, for instance spin waves.

For weak interaction, on the other hand, the system is a metal in the
sense of a Fermi liquid. This
is true if we neglect ordering phenomena, i.e.,
we focus on the paramagnetic phase.
Hence this system is dominated by the quasiparticle excitations.
No collective modes play a major role. 

In the vicinity of the metal-to-insulator transition the Fermi 
liquid must change such that precursors of the collective magnetic
excitations of the insulator emerge. Vice versa, the collective
excitations in the insulator must acquire finite life times so that
they broaden considerably in the corresponding spectra if we pass in the 
insulating regime towards the metallic one. Such behavior can be measured 
by inelastic neutron scattering for example.


Hence it is of particular interest to understand the dynamics of 
the important excitations in the vicinity of an interaction induced
metal-to-insulator transition. In the present work, we focus
on the single-particle excitations, that means we investigate
the single particle propagation. The investigation is done in the
framework of the dynamic mean-field theory (DMFT) which maps the
lattice model to an effective single site problem with a self-consistency
condition.\cite{metzn89a,Mueller-Hartmann-1989b,georg92a,Jarrell-1992} 
The single site is coupled to a bath of non-interacting
fermions which can be represented as a semi-infinite chain.

The semi-infinite chain and its head where the interaction takes place
can be tackled numerically by algorithms for one-dimensional
systems. We will use density-matrix renormalization for this purpose.
\cite{White-1992,White-1993}
Since the dynamic properties have to be determined we use the
dynamic density-matrix renormalization with correction vector.
\cite{Ramasesha-1997,Kuehner-1999} Thereby, we dispose of 
a numerical means which can resolve spectral properties not only
at low energies but also at high energies.\cite{Raas-2004} 
This makes it possible to determine sharp spectral features away from the 
Fermi energy also  quantitatively.\cite{Karski-2005}

The main aim of the present article is to provide comprehensive
spectral data in the above mentioned theoretical framework. 
In particular, we present momentum-resolved single-particle
spectral densities in the insulating and in the metallic regime.

In the following Sec.~\ref{sec:model}, we will present the model under
investigation, i.\,e.~the half-filled one-band Hubbard model 
with semi-elliptic density of states (Bethe lattice) at zero temperature. 
Next, in Sec.~\ref{sec:method}, we will discuss the method.
We use the dynamic mean-field theory (\mbox{DMFT}, Sec.~\ref{sec:dmft}) 
which maps the problem to an effective single-impurity Anderson model
(SIAM, Sec.~\ref{sec:siam})  with a self-consistency condition. The
SIAM is treated numerically by a dynamic density-matrix renormalization 
(\mbox{D-DMRG}) impurity solver
(Sec.~\ref{sec:ddmrg}). The results of this approach are presented in
Sec.~\ref{sec:res}, with the spectral properties in
the paramagnetic \emph{insulating phase} (Sec.~\ref{sec:dyn:ins})
and the paramagnetic \emph{metallic phase} (Sec.~\ref{sec:dyn:met}).
The results reveal insights in the nature of the
Mott-Hubbard metal-to-insulator transition and provide a numerical proof for
previous investigations which were based on the hypothesis of the
\emph{separation of energy scales}. Finally, a  summary will be given in
Sec.~\ref{sec:summary}.


\subsection{Model}
\label{sec:model}

We use the generic model for strongly correlated systems in solids, namely
the Hubbard model.\cite{kanam59,hubba63,gutzw63} It is given by the Hamiltonian
\begin{equation}
  \label{eq:ham_hubb}
  \mathcal{H} =
  U \sum_i \left(n_{i,\uparrow  }-\nicefrac{1}{2}\right)
           \left(n_{i,\downarrow}-\nicefrac{1}{2}\right)
  - t \sum_{\langle i,j\rangle, \sigma}
    c^\dagger_{i,\sigma} c^{\phantom\dagger}_{j,\sigma}
\end{equation}
where $i$, $j$ denote sites on a lattice with $\langle i,j\rangle$ being
nearest neighbors, $\sigma\in\{\uparrow,\downarrow\}$ the spin and
$c^{(\dagger)}_{i;\sigma}$ the electron annihilation (creation) operator and
$n_{i;\sigma}$ their occupation operator.
This model contains the basic ingredients of the problem,
i.e., a local repulsive interaction and a kinetic energy consisting
of hopping from site to site. The interaction is diagonal in real space and
hence tends to make the eigenstates local in real space.
The kinetic energy is diagonal in momentum space and hence tends to make
the eigenstates local in momentum space implying extended states
in real space. So the interaction favors an insulating phase whereas the
kinetic energy favors a conducting, metallic phase. They are the
antagonists and depending on their relative strength the system is
metallic or insulating. 

The transition between these two phases
has been a long-standing issue. By now, evidence emerges that the
transition is continuous at zero temperature 
\cite{kotli99,Karski-2004,Karski-2005} 
whereas it is of first order at any finite temperature.\cite{potth03b} 
The continuous transition at zero temperature is peculiar. It cannot be seen as
 an ordinary second-order transition. It should rather be seen
as a marginal first-order transition with zero hysteresis. At zero temperature,
the free energy is continuously differentiable at the transition
 because the ground-state energy is  continuously differentiable. 
The residual entropy jumps even at zero
temperature because the ground-state of the metal is the non-degenerate
Fermi sea. The insulator, however, is governed by the spin degrees of 
freedom. Since we do not consider any long-range order they remain
free down to zero energy. The freedom to choose the spin locally $\uparrow$ 
or $\downarrow$ in the paramagnetic insulating phase implies a finite residual 
entropy of $\ln 2$ per site. But at zero temperature, the
discontinuous entropy does not imply jumps in the other thermodynamic
quantities.


\section{Method}
\label{sec:method}

In this section we present the details of the methods that
we used to tackle the model and to compute the desired 
single-particle dynamics. The key points are the dynamic
mean-field theory, the single-impurity Anderson model and
the dynamic density-matrix renormalization.


\subsection{Dynamic mean-field theory}
\label{sec:dmft}

The basic idea of the dynamic mean-field theory is to consider 
the lattice problem under study in the limit of infinite coordination
number $z\to \infty$. This means that the lattice is generalized in a way 
that $z$ is a tunable parameter. Very often, this is realized by considering
related lattices in $d$-dimensional space. Then $z\propto d$ and 
the desired limit is the limit of infinite dimensions.

The limit $z\to\infty$ is well-defined only for a particular scaling
of the matrix elements in the Hamiltonian (\ref{eq:ham_hubb}) which link
different sites. In particular, the hopping to the nearest-neighbors must
be scaled like $1/\sqrt{z}$.\cite{metzn89a,Mueller-Hartmann-1989b} Then
the single-particle propagator $G_{ij}$ between site $j$ and site $i$ 
scales like 
\begin{equation}
\label{eq:prop-scale}
G_{ij} \propto z^{||i-j||/2}
\end{equation}
where we use $||i-j||$ for 
the taxi cab metric. This metric counts how many hopping processes are at 
least  necessary to go from $j$ to $i$.

Considering the standard diagrammatic expansion in powers of the interaction
one realizes that the scaling (\ref{eq:prop-scale}) implies that many
diagrams vanish for $z\to\infty$. Only those which are either completely
local or which are sufficiently numerous lead to non-vanishing 
contributions. Closer inspection shows that only those dressed skeleton 
diagrams lead to non-vanishing contributions which are
completely \emph{local}.\cite{metzn89a,Mueller-Hartmann-1989b}
This  is called the collapse of the diagrams because only a single site
occurs. For the proper self-energy $\Sigma$ this implies that
it is local; only  $\Sigma_{ii}$ needs to be considered.

In spite of the very substantial reduction of the number of
relevant diagrams their summation is not directly possible.
Hence the dynamic mean-field theory requires an additional
element. This is the observation that the \emph{same} local dressed skeleton
diagrams occur in the treatment of a single-impurity Anderson model (SIAM) 
where the interaction takes place at the impurity which is coupled to
a fermionic bath.\cite{georg92a,Jarrell-1992,uhrig96a}
Hence, the \emph{same} self-energy is obtained \emph{if} the local dressed 
propagators are the same. Let us use capital letters for the
local dressed propagator $G:=G_{ii}$ and for the local self-energy 
$\Sigma:=\Sigma_{ii}$ in the  lattice model in DMFT. We use small letters for 
the local dressed propagator $g$ and the self-energy $\sigma$ at the impurity
of the effective single-site problem. Then the above statement reads
\begin{subequations}
\label{eq:selfcon}
\begin{equation}
\label{eq:self-energy}
\Sigma(\omega)=\sigma(\omega)
\end{equation}
if
\begin{equation}
\label{eq:local-propagator}
G(\omega)=g(\omega)
\end{equation}
\end{subequations}
holds. Note that this does \emph{not} imply that $G^0=g^0$ where
we use the superscript $^0$ for the bare, undressed propagators.

For completeness, we mention that the self-consistency
equations \eqref{eq:selfcon}, which we derived from the expansion
in dressed skeleton diagrams, can also be obtained from the action
by the so-called cavity argument.\cite{Georges-1996} A derivation
in the strong coupling limit is given in the review 
\onlinecite{Georges-1996}.
An elegant derivation of the DMFT including the possibility to generalize 
from single site to cluster systems has  been found by Potthoff \cite{potth03b}
in the framework of variational  self-energy functionals.

Eqs.~\eqref{eq:selfcon} face us with the problem that we do not know the
bare propagator of the SIAM. This amounts up to a generic
self-consistency problem where we first have to guess a $g^0$ and
then we can verify whether the conditions 
\eqref{eq:selfcon} are fulfilled.

Following this roadmap, two more fundamental relations are needed,
namely the Dyson equations of the lattice problem and of the SIAM.
The local self-energy $\Sigma_{ii}$ acts on every site equally since
we consider the uniform, paramagnetic phase. For this reason we could
omit the subscript and pass on to $\Sigma$. This spatially constant energy
acts like a global energy shift so that it can be accounted for by
\begin{equation}
\label{eq:dyson-lattice}
G(\omega)=G^0(\omega-\Sigma(\omega)),
\end{equation}
which is the Dyson equation for the lattice problem.

In the single-site problem, the self-energy $\sigma(\omega)$ acts
only on the impurity site. The bare propagator from the impurity site
to the impurity site is $g^0(\omega)$ so that the Dyson equation is
given by the geometric series
\begin{eqnarray}
\nonumber
g(\omega) &=&  g^0 + g^0\sigma g^0
+g^0\sigma g^0\sigma g^0+\ldots+g^0(\sigma g^0)^n+\ldots
\\
&=& \frac{g^0(\omega)}{1-g^0(\omega)\sigma(\omega)}.
\end{eqnarray}
This relation can also be written as
\begin{equation}
\label{eq:dyson-siam}
\nicefrac{1}{g} = \nicefrac{1}{g^0}-\sigma.
\end{equation}

\begin{figure}[ht]
  \includegraphics[width=0.98\columnwidth]{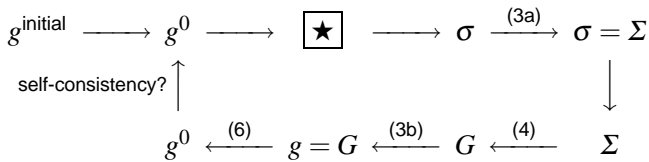}
  \caption{\label{fig:sc:gen}
    Schematic representation of the iterative self-consistency cycle for an
    arbitrary lattice. In the step marked with \ding{72} the dynamics of an
    effective single-impurity Anderson model is calculated by a \mbox{D-DMRG}
    impurity solver.
  }
\end{figure}

The above equations are used to set up an iteration cycle to
find a self-consistent solution. It is illustrated in Fig.\ \ref{fig:sc:gen}.
Starting in the upper left corner, an initial guess is used to 
define the SIAM which is then solved by an impurity solver providing $\sigma$.
Then $\sigma$ is used as lattice self-energy $\Sigma$ providing
via the lattice Dyson equation (\ref{eq:dyson-lattice}) the dressed
propagator for the lattice. By the condition (\ref{eq:local-propagator})
the dressed propagator of the SIAM is known which in turn yields 
via (\ref{eq:dyson-siam}) its bare counterpart so that we are at the 
beginning of the cycle again. If the bare propagator is close enough
to the previous assumption the iterations are stopped. Details of the
precise meaning of how close is close enough are given in Sec.\
\ref{sssec:sc}.

The above sketched cycle can be used for any lattice, i.e.,
for any $G^0$. For a particular $G^0$ with semi-elliptic spectral density
 (DOS) $\rho^0(\omega) :=\nicefrac{-1}{\pi}\mathfrak{Im}G^0(\omega)$
\begin{equation}
  \label{eq:bethe_dos}
  \rho^0(\omega) = \frac{2}{\pi D^{2}}\sqrt{D^{2}-\omega^{2}},
\end{equation}
the cycle can be simplified considerably. A semi-elliptic spectral
density $\rho^0(\omega)$ 
corresponds to the so-called Bethe lattice with infinite
branching ratio.\cite{econo79} Note that we do not use any other
feature from the Bethe lattice other than \eqref{eq:bethe_dos}.
Hence our calculation can be viewed as treating a translationally
invariant lattice with a semi-elliptic DOS.
This is a good starting
point for generic spectral densities since it is bounded and it possesses
square-root singularities at the band edges like three-dimensional 
densities-of-states have. Otherwise it is featureless.

The key feature of $G^0$ with semi-elliptic
$\rho^0(\omega)$ is its particularly simple
continued fraction representation with constant coefficients
\begin{equation}
  \label{eq:bethe_contfrac}
G^0(\omega) = 
  \frac{1}{\omega-
    {\displaystyle\frac{\nicefrac{D^2}{4}}{\omega-
        {\displaystyle\frac{\nicefrac{D^2}{4}}{\omega-\cdots}}}}}.
\end{equation}
In order to use this fact for simplification we write the bare propagator
of the SIAM with the help of the so-called hybridization function
$\Gamma(\omega)$ as
\begin{equation}
g^0(\omega) = \frac{1}{\omega-\Gamma(\omega)}
\end{equation}
where the continued fraction of $\Gamma(\omega)$ shall be 
parametrized like
\begin{equation}
  \label{eq:hyb_cf}
  \Gamma(\omega) = 
  \frac{V^2}{\omega-
    {\displaystyle\frac{\gamma_0^2}{\omega-
        {\displaystyle\frac{\gamma_1^2}{\omega-\cdots}}}}}.
\end{equation}
Then, based on Eqs.\ (\ref{eq:dyson-siam},\ref{eq:self-energy}) 
the dressed propagator of the SIAM reads
\begin{equation}
\frac{1}{g(\omega)} = \omega -\Gamma(\omega) - \Sigma(\omega).
\label{eq:compare1}
\end{equation}
By Eqs.\ (\ref{eq:dyson-lattice},\ref{eq:bethe_contfrac}) the dressed  
propagator of the lattice can be expressed to be
\begin{eqnarray}
\nonumber
\frac{1}{G(\omega)} &=& \omega -\Sigma(\omega) - 
 \frac{\nicefrac{D^2}{4}}{\omega-\Sigma -
    {\displaystyle\frac{\nicefrac{D^2}{4}}{\omega-\Sigma -
        {\displaystyle\frac{\nicefrac{D^2}{4}}{\omega-\Sigma\cdots}}}}}
\\
 &=& \omega -\Sigma(\omega) - \left(\nicefrac{D^2}{4} \right) G(\omega),
\label{eq:compare2}
\end{eqnarray}
where we exploited that the continued fraction does not change
when it is evaluated at a deeper level because its coefficients 
are constant.

Based on the self-consistency condition (\ref{eq:local-propagator}) we
set Eqs.\ (\ref{eq:compare1},\ref{eq:compare2}) equal and obtain the simpler 
self-consistency condition
\begin{equation}
  \label{eq:hyb_green_sc}
  \Gamma(\omega) = \frac{D^{2}}{4}G(\omega).
\end{equation}
This equation is more easily used because it provides a direct way to compute
the hybridization function $\Gamma(\omega)$ of the next iteration of
the SIAM from the dressed lattice propagator $G$. No explicit computation of
intermediate self-energies is needed which can be severely hampered by
numerical errors, see below. A first conclusion which can be drawn from
(\ref{eq:hyb_green_sc}) is that $V=D/2$.

\begin{figure}[ht]
  \includegraphics[width=0.98\columnwidth]{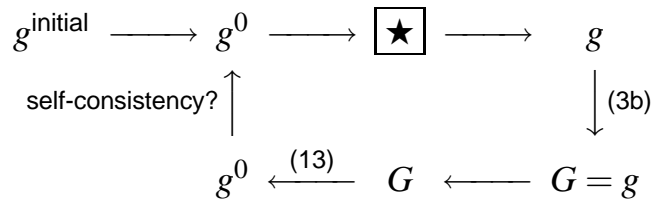}
  \caption{\label{fig:sc:bethe}
    Schematic representation of the iterative self-consistency cycle for 
    semi-elliptic density of states (DOS) with nearest-neighbor hopping. 
    The step \ding{72} stands for
    the \mbox{D-DMRG} {impurity solver}.
  }
\end{figure}

The resulting cycle is given graphically in
Fig.\ \ref{fig:sc:bethe}. This form of the iteration cycle is used
in the present work.


\subsection{Single-Impurity Anderson Model}
\label{sec:siam}

There are various ways to set up the effective single-impurity problem
because the precise topology of the fermionic bath, to which the impurity
couples, does not matter. As we have seen in the equations above we
are only concerned with the local quantities on the impurity. Hence the
only relevant property of the fermionic bath is the bare propagator
$g^0$ or the hybridization function $\Gamma$, respectively.

\begin{figure}[ht]
  \includegraphics[width=0.98\columnwidth]{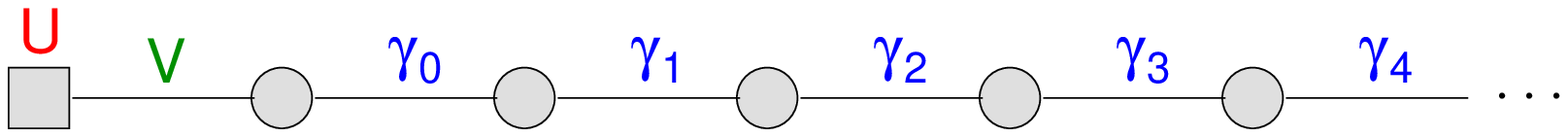}
  \caption{\label{fig:chain-bath}
    (Color online) Sketch of the semi-infinite chain which is used as topology 
    for the single-impurity Anderson model.
  }
\end{figure}

Because we aim at the application of density-matrix renormalization,
which has been developed for one-dimensional systems in particular,
we favor the representation of the bath as semi-infinite chain, see
Fig.\ \ref{fig:chain-bath}.
This is no restriction since mathematically any hybridization function which 
has a non-negative spectral density can be represented as continued
fraction \cite{Pettifor-1985} as given in Eq.\ (\ref{eq:hyb_cf}). By complete
induction it is straightforward to see that the choice of the hopping elements
shown in Fig.\ \ref{fig:chain-bath} implies Eq.\ (\ref{eq:hyb_cf}).

The  Hamiltonian of the SIAM represented as linear chain
reads\cite{Hewson-1993}
\begin{gather}
  \notag
  \mathcal{H} = U
  \left(n_{d,\downarrow}-\nicefrac{1}{2}\right)
  \left(n_{d,\uparrow  }-\nicefrac{1}{2}\right)
  + V
  \sum_\sigma\left(
    d_{\sigma}^{\dagger} c_{0,\sigma}^{\phantom{\dagger}} + \text{h.c.}
  \right)\\
  \label{eq:ham_siam}
  + \sum_{n=0,\,\sigma}^\infty \gamma_{n}
  \left(
    c_{n,\sigma}^\dag c_{n+1,\sigma}^{\phantom{\dagger}} + \text{h.c.}
  \right) .
\end{gather}
The correlations stem from the density-density coupling on the impurity site
($d^{(\dagger)}$ operators) at the head of a semi-infinite chain of free
fermions ($c^{(\dagger)}$ operators). For comprehensive information on
the physics of single-impurity Anderson models the reader is referred
to Ref.~\onlinecite{Hewson-1993}.


The determination of the local propagator $g(\omega)$ in the effective SIAM
\eqref{eq:ham_siam} for a given hybridization function $\Gamma(\omega)$
is the most difficult part in the self-consistency cycles 
(cf.~Figs.~\ref{fig:sc:gen} and \ref{fig:sc:bethe}). The necessary
tool is called the \emph{impurity solver}.

The dynamics one has to compute is the dynamics of
the fermionic single-particle propagator of the impurity electron. Aiming at
the properties at $T=0$ it reads
\begin{align}
  \label{eq:green_siam}
  \begin{split}
    g(\omega+\mathrm{i}\eta) = &\left\langle 0\left|
        d_\sigma
        \frac{1}{\omega+\mathrm{i}\eta-(\mathcal{H}-E_0)}
        d^\dagger_\sigma
      \right|0\right\rangle \\
    + &\left\langle 0\left|
        d^\dagger_\sigma
        \frac{1}{\omega+\mathrm{i}\eta+(\mathcal{H}-E_0)}
        d_\sigma
      \right|0\right\rangle\ .
  \end{split}
\end{align}
Here the ground-state is denoted by $|0\rangle$ and its energy by $E_0$. Since
we aim at the paramagnetic phase the propagator has no dependence on the
spin index $\sigma$ and we do not denote it. The
frequencies $\omega$ and $\eta$ are real. The standard retarded propagator
is retrieved for $\eta\to0+$
\begin{equation}
  g_\text{R}(\omega) = \lim_{\eta\to0+}g(\omega+\mathrm{i}\eta).
\end{equation}
The wanted quantity is the spectral density
$\rho(\omega):=-\pi^{-1}\mathfrak{Im}\,g_\text{R}(\omega)$. If necessary the
real part of $g_\text{R}(\omega)$ can be obtained from $\rho(\omega)$
by the Kramers-Kronig relation
\begin{equation}
  \label{eq:kkt_re}
  \mathfrak{Re}\,g_\text{R}(\omega) =
  \mathcal{P}\int\limits_{-\infty}^{\infty}
  \frac{\rho(\omega')}{\omega-\omega'}\,\mathrm{d}\omega' ,
\end{equation}
where $\mathcal{P}$ denotes the principal value.


\subsection{Dynamic Density-Matrix Renormalization}
\label{sec:ddmrg}

\subsubsection{Algorithm}
Because there are essentially no analytic approaches available which solve the 
general SIAM, reliable numerical  approaches for dynamic correlations are very 
important. Our choice is the dynamic density-matrix renormalization which 
is an excellent tool for calculations for open one-dimensional systems
at zero temperature. 

To obtain $g(\omega)$ for a given $\Gamma(\omega)$ we use a combination of the
dynamic density-matrix
renormalization\cite{Hallberg-1995,Ramasesha-1997,Kuehner-1999} (\mbox{D-DMRG})
and the least-bias deconvolution algorithm.\cite{Raas-2005a} The \mbox{D-DMRG}
is the generalization of the standard density-matrix renormalization
group\cite{White-1992,White-1993} (DMRG) for the calculation of dynamic
quantities. A key advantage over other $T=0$ approaches like the numerical 
renormalization group is that the resolution can be high not only at low 
energies but also at high energies.\cite{Raas-2004} 
We implemented the \mbox{D-DMRG} in a correction vector (CV)
scheme\cite{Ramasesha-1997,Kuehner-1999,Hoevelborn-2000} with optimized direct
matrix inversion, for details see Refs.~\onlinecite{Raas-2004} and
\onlinecite{Raas-2005b}.

The actual calculations are not performed for the fermionic
representation \eqref{eq:ham_siam} but for a spin representation
involving two semi-infinite spin chains which are coupled
at their heads.
This spin representation replaces each fermionic level with states
empty and occupied by the two spin states $\uparrow$ and $\downarrow$.
The change of basis between both representations is accomplished by
two Jordan-Wigner transformations, one for the $\uparrow$ electrons and
one for the $\downarrow$ electrons, see  Ref.~\onlinecite{Raas-2004}.
The resolvents appearing in Eq.~\eqref{eq:green_siam}
are expressed as resolvents of the equivalent spin system involving
spin flips at the head of the chains. The advantage for the DMRG approach is 
that in the spin chain description each site carries only a two dimensional 
Hilbert space instead of a four dimensional one. 

The correction vector \mbox{D-DMRG} provides data points at given frequencies
$\omega=\omega_i$ for finite values of $\eta=\eta_i$
\begin{subequations}
\begin{eqnarray}
  \label{eq:convolv0}
   g_i &:=& -\frac{1}{\pi}\mathfrak{Im}\,g(\omega_i+\mathrm{i}\eta_i)
   \\ 
   &=& \rho(\omega) \otimes L_{\eta_i}(\omega)\Big|_{\omega=\omega_i}
   \\
   &=& \frac{1}{\pi}\int\limits_{-\infty}^\infty 
   \frac{\eta_i\rho(\omega)\mathrm{d}\omega}{(\omega-\omega_i)^2+\eta_i^2},
\end{eqnarray}
\end{subequations}
where $ \otimes L_{\eta_i}(\omega)$ stands for the convolution with
a Lorentzian of width $\eta_i$. No data can be
obtained directly at $\eta=0$ since the inversion of the Hamiltonian is
singular and cannot be achieved numerically in a stable way. 
Furthermore, we cannot numerically treat an infinite chain but only a finite
one.

Henceforth, we will call data $\{g_i\}$  \emph{raw} data because it is obtained
at finite values of $\eta_i$. Apart from the constraint from matrix inversion, 
a sufficiently large broadening $\eta_i$ is used in order to calculate 
continuous spectral densities  resembling the spectral densities 
in the thermodynamic limit. The finite-size effects are washed out.
In a second step, we aim at retrieving the unbroadened spectral density
$\rho(\omega)$ as well as possible.

We use the \mbox{D-DMRG} raw data calculated for a finite set of frequencies
$\omega_i$ with non-zero broadenings $\eta_i$ as input of a deconvolution
scheme to extract the information on the spectral density $\rho(\omega)$,
i.\,e.~the relevant dynamic properties of the infinite system. The
deconvolution scheme of our choice is the least-bias (LB) approach
\cite{Raas-2005a} which belongs to the class of maximum entropy
methods.\cite{Press-1992}  The combination of \mbox{D-DMRG} and LB
deconvolution has already proven itself in the investigation of the high-energy
dynamics of the SIAM by providing a well-controlled resolution for all
energies.\cite{Raas-2005a,Raas-2005b}  By construction, the LB ansatz yields  a
positive and continuous result for the density of states $\rho(\omega)$ (DOS). 
The positiveness assures the causality of the solution, which  is necessary
for calculating the continued fraction, see App.\ \ref{app:cf}, whereby
 the DMFT self-consistency cycle is closed. By using
the LB deconvolution we avoid any kind of arbitrariness introduced
by the choice of a discretization mesh.

Of course, the direct computation of a continuous density from data obtained 
for a finite system involves an approximation. In practice, one must
pay attention whether the system size and the broadening considered make 
it possible to deduce continuous densities, see also below.

Nishimoto and co-workers \cite{Nishimoto-2004a,nishi06}
advocate a different approach which has been baptized
``Fixed-Energy'' DMFT where the problem of deducing \emph{densities}
is circumvented by discretizing the continuous frequency in a number
of bins. For these bins the spectral \emph{weights} are computed. 
On the one hand, spectral weights are better behaved for finite system sizes.
But on the other hand, the DMFT self-consistency holds only for the
continuous quantities in Eq.~\eqref{eq:selfcon}. So any extension
to discrete quantities introduces some ambiguity.\cite{Eastwood-2003}
Balancing these two contrary aspects we have chosen to work with
the continuous spectral densities as obtained from the LB ansatz.

In the framework of the \mbox{D-DMRG} Jeckelmann has formulated a variational
principle for the wanted dynamic correlation.\cite{Jeckelmann-2002} A certain 
functional is minimum for the correct dynamic correlation. Its value 
yields the $g_i$ as defined in \eqref{eq:convolv0}. Due to the minimum 
principle this approach is rather robust. Its main advantage is that even if
the numerical dynamic correlation possesses an error $\delta$ the 
values $g_i$ will be exact up to quadratic order $\delta^2$.
From the algorithmic view point the minimization required for
Jeckelmann's approach is more demanding than the matrix inversion in the 
D-DMRG with correction vector. Hence we stick to an optimized matrix inversion.
\cite{Raas-2004,Raas-2005b}
In our opinion, the optimum strategy is to search for the best correction
vector by optimized matrix inversion and then to use the variational
functional for the evaluation of the $g_i$.

\subsubsection{Deconvolution and Finite-Size Effects}

The choice of the $\eta_i$ for a given set of $\omega_i$ is restricted by the
considered chain length $L$.
If the $\eta_i$ are chosen too small, the extracted spectral properties reveal
too strong finite-size effects. If the $\eta_i$ are chosen too large,
essential features in the spectral properties cannot be resolved properly and
therefore remain smeared out, even after the deconvolution. 
Thus, the optimum value of each $\eta_i$ has to been chosen with care 
depending on the width of the features which have to be resolved. 
In our approach the optimum choice is characterized by
\begin{equation}
  \label{eq:fsites_broadening}
  \eta_i\approx\Delta\omega_i \gg W^{*}/L
\end{equation}
where $\Delta\omega_i$ denotes the distance between two neighboring 
frequencies and $W^{*}$ is the bandwidth. The approximate equality 
between $\eta_i$ and $\Delta\omega_i$ reflects the fact that it is
not reasonable to numerically measure the spectral density which is 
broadened by $\eta_i$ at frequencies much closer than  $\eta_i$.
This does not lead to an increase of relevant information about the 
underlying unbroadened spectral density. In contrast, it might happen
that small errors lead to inconsistencies of the too closely positioned
data points so that the subsequent deconvolution leads to strongly
oscillatory results.\cite{Raas-2005a}

The inequality $\Delta\omega_i \gg W^{*}/L$ in \eqref{eq:fsites_broadening}
results from the approximation of the infinite system under study
by a finite system. If the points $\omega_i$, at which the broadened
spectral density is measured numerically, were too closely spaced
the finite-size effects would be discernible. 
Hence the raw data $\{g_i\}$ could not 
be processed as if it resulted from a continuous spectral density.

Finally, we point out, that apart from the 
resolution aspect, the choice of $\eta_i$ influences the stability and the 
performance of the correction vector \mbox{D-DMRG}. 
For very small values of $\eta_i$ the matrix inversions are
numerically very demanding and time-consuming.

\subsubsection{Continuous Fraction}

The LB algorithm and the Kramers-Kronig transformation allow us to extract the
full unbroadened Green function $g(\omega)$ to good accuracy
from the raw \mbox{D-DMRG} data for an arbitrary $\omega$. 
Thus we can have an arbitrary number of supporting points for
further numerical calculations. The hopping elements $\gamma_n$ of the 
effective SIAM \eqref{eq:ham_siam} for the next iteration in the 
self-consistency cycle are calculated recursively by a continued fraction
expansion.\cite{Pettifor-1985,Viswanath-1994}
Note, that even  tiny regions of small negative spectral density
spoil the continued fraction expansion completely. The details
of the calculation of the continued fraction are given in App.~\ref{app:cf}.

\subsubsection{Numerical Criteria for Self-Consistency}
\label{sssec:sc}

We have realized the iterative cycle shown in Fig.\ \ref{fig:sc:bethe}.
Since we perform the calculation of the spectral densities numerically,
including an advanced deconvolution scheme, the calculation of two 
\emph{identical}
spectral densities within the iterative self-consistency cycle is
not possible by principle. Consequently, a certain error tolerance has to be
included to the termination conditions.

A solution within our DMFT self-consistency cycle is understood as
self-consistent, if at least the following conditions are fulfilled:
\begin{enumerate}
\item Two deconvolved spectral densities calculated consecutively within the
  self-consistency cycle obey for all $\omega$,
\begin{equation}
  |\rho^{(i)}(\omega)-\rho^{(i+1)}(\omega)|<\varepsilon\ ,
\end{equation}
where $\varepsilon=4\times 10^{-3}/D$ for the metallic and
$\varepsilon=10^{-3}/D$ for the insulating solutions.
\item The average double occupancy and the ground-state energy per lattice site
  calculated in succession within the self-consistency cycle are stable. That
  means that they differ by less than $10^{-2}\%$.
\end{enumerate}
For the insulating solutions, we additionally require that the \emph{static}
single-particle gaps calculated consecutively by means of the standard DMRG 
algorithm differ by
less then $1\%$. The higher value of $\varepsilon$ for the metallic
solutions has to be chosen because the metallic spectral densities 
display a high complexity including many sharp features. In contrast, the
spectral densities of the insulating solutions are rather featureless
and thus converge more easily.

We stress that the upper conditions are the  minimum
requirement and that most of our solutions surpass them considerably.
But  we observed that it is sufficient to terminate the
self-consistency cycle once the upper conditions are fulfilled.
The physical quantities obtained from the continued fractions computed in two 
consecutive iterations are almost identical. Further iterations yield no 
further improvement. Moreover, we have confirmed that our final
continued fraction coefficients do not display persistent trends of change
if iterated further. They rather display very small oscillations around
the results which we consider to be converged.

\subsubsection{Numerical Stability of the Metallic and the Insulating Solution}
\label{sssc:numer-stabl}

There is a parameter region in the Hubbard model in infinite dimensions
with an interaction $U$ between $U_\text{c1}$ and $U_\text{c2}$ where both 
solutions are locally stable. In a numerical treatment of this situation one 
has to introduce a bias, preferably minute, which selects the solution we
want to find. 

This can be done elegantly by exploiting an even-odd effect.
If we consider only the impurity site, the bare DOS $\rho^0(\omega)$ 
consists of a single $\delta$-peak at $\omega=0$. Any arbitrary finite
repulsion $U>0$ splits this peak into two so that  a gap $\Delta=U/2$
occurs. The corresponding self-energy has a pole at $\omega=0$ which
is necessary to induce the splitting of the central peak. Hence this
self-energy is the self-energy of an insulator.

Any SIAM with an odd number of fermionic sites (one impurity site
and an even number of bath sites) has a bare DOS $\rho^0(\omega)$ with 
a $\delta$-peak at $\omega=0$. This lowest energy level can be
occupied or unoccupied without changing the ground-state energy.
A finite repulsive interaction will lift this degeneracy partly and
split the central  $\delta$-peak in two and a gap occurs.
Of course, if the levels in the bare system lie very
close to each other the occurring gap can be very small.
But in any case, the solution resembles an insulating one.
We conclude that the use of a SIAM with an odd number of sites
implies a small bias towards an insulating solution. Obviously,
the relative bias is of the order of the inverse system size $1/L$.

Conversely, considering a SIAM with an even number of sites amounts up to a
small bias towards a self-energy belonging to a metallic phase.
The imaginary part of $\Sigma$ at $\omega=0$ is zero.

While the metallic solutions correspond to the Fermi sea as
unique ground-state the paramagnetic insulating solution implies
a macroscopically large degeneracy in the lattice problem.
In the SIAM, the degeneracy is only two. It stems from the
fact that the spin of the impurity is not completely screened
by the coupling to the bath because the bath is gapped by $\Delta$.
So there is no DOS at low energies and the renormalization of the
unscreened moment stops at the lower cut-off $\Delta$.
Indeed, we found that the insulating baths
show a two-fold degenerate ground-state or 
two almost degenerate low-lying states. 

The spin degeneracy in the insulator poses a problem to the numerical
treatment. It happens easily that the DMRG algorithm looking for
the ground-state provides a state with a certain finite magnetization
in $z$-direction. If such a solution is used further in the 
iteration cycle results are produced which do not describe the
paramagnetic insulating phase. In order to avoid this problem,
we first search for the ground-state  $|1\rangle$ with energy $E_1$
using a standard Lanczos or Davidson algorithm starting from an initial guess 
$|I_1\rangle$. The degenerate or almost degenerate second state $|2\rangle$ is 
found from a second run starting from the orthogonal initial guess 
$|I_2\rangle =  |I_1\rangle - \langle 1|I_1\rangle |1\rangle$. 
With both states we can construct a ground-state without 
magnetization. We use the normalized superposition
$|0\rangle = \alpha_1|1\rangle + \alpha_2|2\rangle$
with $1=|\alpha_1|^2+|\alpha_2|^2$ where the coefficients
are chosen such that 
\begin{equation}
  \label{eq:halffilling}
  \left\langle0\left|
      d_{\sigma}^{\phantom{\dagger}}d_{\sigma}^{\dagger}\right
  |0\right\rangle = \nicefrac{1}{2}.
\end{equation}
for $\sigma=\{\uparrow,\downarrow\}$ which ensures the absence of a magnetization
in $z$-direction.

Finally, we note that the treatment of the spectral density in the
region of the single-particle gap
$\Delta$ in the insulating phase requires care. The LB deconvolution
scheme does not allow for zero spectral density so that some artifacts
need to be removed before the 
continued fraction expansion can be calculated as explained in 
Appendix \ref{app:cf}. The details of the removal 
of deconvolution artifacts are given in App.\ \ref{app:gap}.


\section{Results}
\label{sec:res}

Here we present the results for various spectral densities
which we have obtained by the approach described so far.
First we focus on the insulating solutions; then we pass on to the metallic
solutions. The results are interpreted as well as compared to previous results.

\subsection{Insulator}
\label{sec:dyn:ins}

\subsubsection{Local Spectral Densities}
\label{sssec:ins:lsd}

The computation of the insulating spectral densities is performed for $121$
fermionic sites, including the impurity. Larger system sizes provide identical
solutions (not shown). This is related to the relatively low complexity of the 
insulating solutions which do not display particular features besides
the insulating gap.

We calculate the raw \mbox{D-DMRG} data on a grid using mostly 
$\Delta\omega_i=\eta_i=0.1D$. Close to the edges of the Hubbard
bands we increase the resolution by choosing $\Delta\omega_i=\eta_i=0.05D$ and
partially $0.025D$ or even $\Delta\omega_i=\eta_i=0.01D$ for $U=2.4D$.
This is done to resolve the gap and the band edges properly. 
Note, that such a small broadening is useful here, although 
$\eta_i=0.01D$ does not obey the inequality \eqref{eq:fsites_broadening},
because it helps to decide where the spectral density is finite or zero.

\begin{figure}[ht]
  \centering\includegraphics[width=\columnwidth]{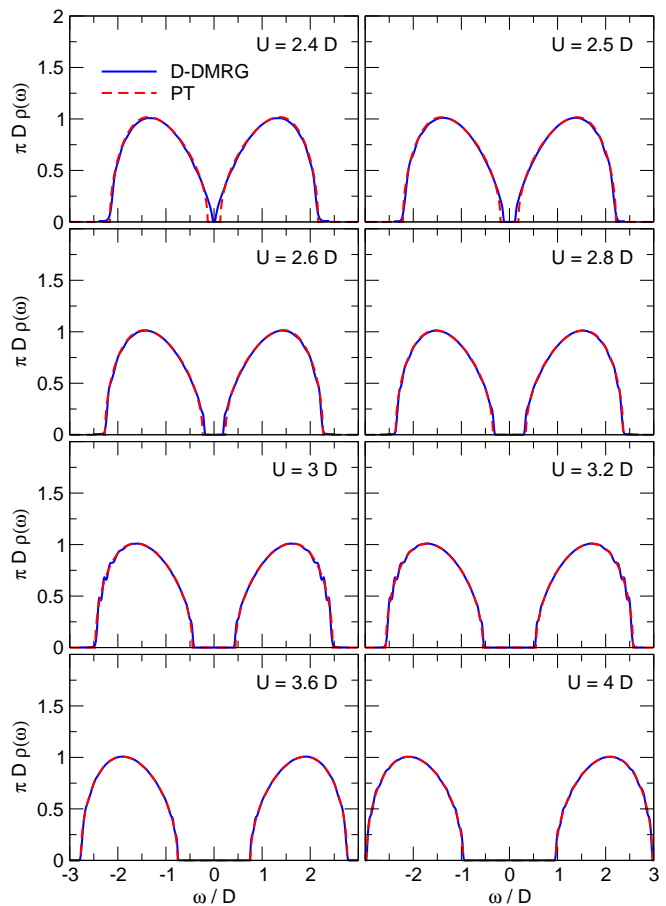}
  \caption{\label{fig:iso_dens}
    (Color online) 
    Spectral densities $\rho(\omega)$
    in the insulating regime (solid line) calculated with a chain of
    $121$ fermions. The dashed lines (almost coinciding) 
    show the results from strong coupling
    perturbation theory.\cite{Gebhard-2003}}
\end{figure} 

\begin{figure}[ht]
  \centering\includegraphics[width=\columnwidth]{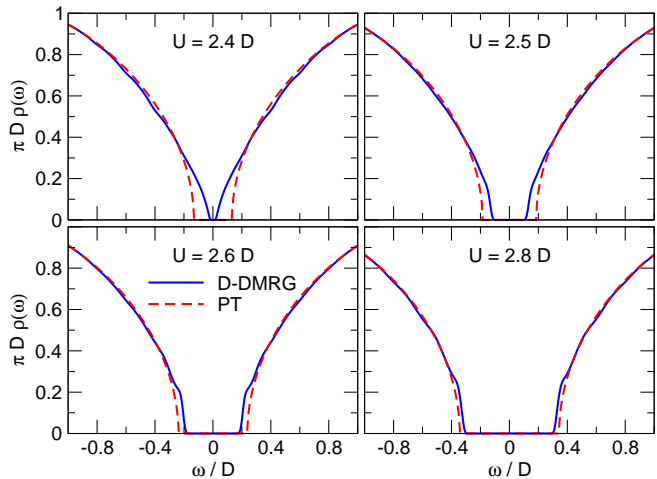}
  \caption{\label{fig:iso_dens_det}
    (Color online) 
    Data from Fig.~\ref{fig:iso_dens} on a zoomed scale close to the inner band
    edges.}
\end{figure}

The insulating solutions in Fig.~\ref{fig:iso_dens} clearly show the lower and
the upper Hubbard band each with a band width of nearly $2D$. The bands are
separated by a well pronounced gap of $2\Delta$.\footnote{Note, that we
use here the definition that $\Delta$ is the energy from the Fermi energy
at $\omega=0$ to the lower inner band edge. This implies that the present 
definition is by a factor of 2 lower than the one in Ref.\ 
\onlinecite{Gebhard-2003}.} 
The center of each Hubbard band is 
located at roughly $\omega=\pm\nicefrac{U}{2}$. 
The shape of the density in the Hubbard bands in our insulating solutions 
is essentially smooth and reveals no significant features. They agree very well
with the perturbative results by Gebhard 
et al.\  \cite{Gebhard-2003} which are based on a strong coupling
expansion in $\nicefrac{1}{U}$. Only for very small gaps $\Delta\approx
0.2D$ a noticeable deviation is discernible, see Fig.\ \ref{fig:iso_dens_det}.
Clearly, these deviations result from the limited number of
terms which are available in the perturbation series.

In Ref.~\onlinecite{Garcia-2004} a DMFT calculation based
on another DMRG impurity solver, the Lanczos method, showed many
substructures in the Hubbard bands. The comparison to our featureless
bands leads us to the conclusion that the peaky substructures in
 Ref.~\onlinecite{Garcia-2004} stem from finite-size effects.
First, such effects enter because only 45 fermionic sites were
considered. Second, the reliable application of the Lanczos impurity solver
would require a large number ($\approx 100$) of target states which cannot 
be handled by the DMRG.

The results in Ref.~\onlinecite{Nishimoto-2004b} differ from ours only
in one important aspect. Nishimoto et al.\ found an upturn of the
spectral density $\rho(\omega)$ close to the inner band edge for
small gaps which is completely absent in our data.
We believe that this difference stems from the different treatment of the
spectral density in the gap region, for our approach see 
Appendix \ref{app:gap}.
In our calculations, Nishimoto's approach to use the extrapolated static gap 
as a cutoff for the spectral density 
did not lead to stable self-consistent solutions.
On iteration, the spectral density becomes peakier and peakier.
The normalization does not hold anymore.

 But we found that
a slightly too large gap value tends to produce features similar to
the upturns in Ref.~\onlinecite{Nishimoto-2004b}.
Our approach is corroborated further by the excellent agreement
between our value of the interaction $U_{\text{c1}}=(2.38\pm 0.02)D$ 
below which no insulating solution exists and the one obtained by extrapolating
the $\nicefrac{1}{U}$ expansion of the ground-state energy. The latter
is supported also by quantum Monte Carlo results.\cite{Bluemer-2005b,blume05a}
In spite of the evidence mentioned above, one may certainly argue
in favor of the Fixed-Energy approach used in 
Ref.~\onlinecite{Nishimoto-2004b} or in favor of the approach used here
because both numerical approaches comprise some approximations. For
further comparison, see also the next subsection Sec.\ \ref{sssec:ins:gap}.

The perturbation theory\cite{Gebhard-2003} implies a square-root behavior
at the band edges. The extremely good agreement between the perturbative and
the numerical results in Figs.\ \ref{fig:iso_dens} and \ref{fig:iso_dens_det}
suggests that a square-root onset is the correct description
\begin{equation}
  \rho_{\text{UHB}}(\omega) \propto (\omega-\Delta)^{1/2}
\end{equation}
at the inner band edges of the upper Hubbard band (UHB).
 
\begin{figure}[ht]
  \centering\includegraphics[width=\columnwidth]{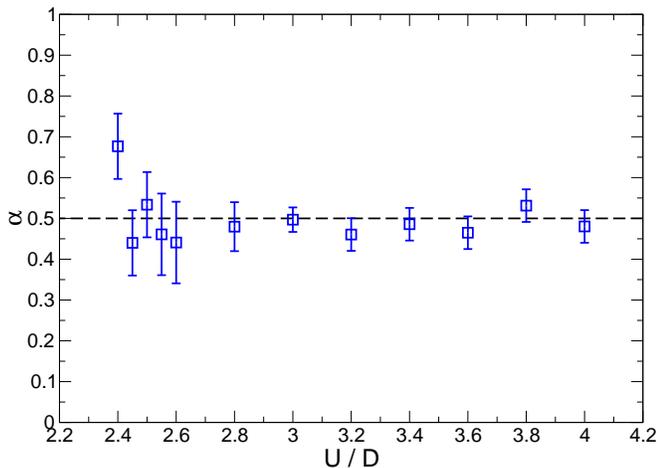}
  \caption{\label{fig:alpha}
    (Color online) 
    Leading exponent $\alpha$ in the behavior of the spectral density
    $\rho(\omega)$ near the band edges ($\omega\to\Delta$) for various values
    of $U$. The dashed line marks the value $\alpha=0.5$ suggested by 
    perturbation theory.\cite{Gebhard-2003}}
\end{figure}

In order to investigate the behavior of the spectral
density quantitatively near the band edges, see Fig.~\ref{fig:iso_dens_det}, 
we fit
\begin{equation}
 \label{eq:fit_alpha_delta}
  f(\omega) =
  \Theta(\omega-\Delta)
  \left[(\omega-\Delta)^{\alpha} P_{3}(\omega-\Delta)\right]
\end{equation}
to the inner side of the upper Hubbard band (not shown). 
Here $\Theta(x)$ denotes the
Heaviside function, $P_n(x)$ is an $n$th order polynomial
\begin{equation}
  \label{eq:fit_alpha_poly}
  P_n(x) = \sum\limits_{i=0}^n c_i x^i\,
\end{equation}
and $\alpha$, $\Delta$, and the $c_i$ are the fit parameters. The parameter
$\alpha$ represents the leading exponent of the power law onset.
Note, that a perfect square-root onset cannot be extracted
using the LB deconvolution scheme since the ansatz for the 
unbroadened data is built from exponential functions.\cite{Raas-2005a}
Hence the fit results depend on a number of assumptions, for instance
the degree $n$ of $P_n(x)$ and the interval in which the least-square
fit is done. Fig.~\ref{fig:alpha} summarizes our results; 
the ambiguities are accounted for by the error bars.
Our results excellently confirm  the square-root onset. The agreement is
very good for large values of the interaction, i.e., deep in the 
insulating regime. Closer to the closure of the gap the uncertainties
grow.

The outer band edges of the Hubbard bands do not show any sharp decay. Instead,
they reveal a smooth drop for $|\omega|\to\infty$ which becomes even smoother 
as $U$ decreases, similar to the behavior of the metallic
solutions, cf.~Sec.~\ref{sec:dyn:met}.

\subsubsection{Insulating Gap}
\label{sssec:ins:gap}

The gap $\Delta$ is the characteristic quantity of the insulator.
Because of the impossibility to determine it directly from the LB 
deconvolved spectral density we use the value which is found
from the fit \eqref{eq:fit_alpha_delta} with fixed exponent
$\alpha=1/2$. We stress that this 
determination of the gap $\Delta$ does not enter the iterations
of the self-consistency cycle,
 except for the removal of artifacts, see App.\ \ref{sapp:inner}.
It is done only at the end of
the iterations once the solution can be considered to be self-consistent
according to the criteria in Sec.\  \ref{sssec:sc}.
This is in contrast to the Fixed-Energy procedure chosen in Ref.\
\onlinecite{Nishimoto-2004b} where the gap enters in the self-consistency
explicitly. Again, it is a priori not clear which approach is 
more advantageous. All we found was that we could not
reach stable self-consistency when using extrapolated gaps in
the devonvolution.

\begin{figure}[ht]
  \centering\includegraphics[width=\columnwidth]{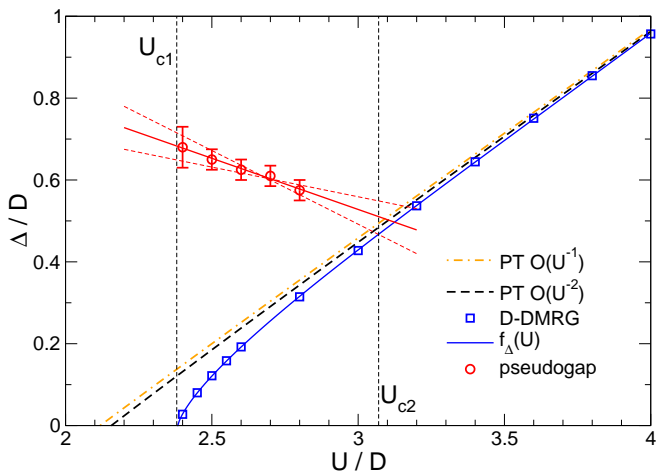}
  \caption{\label{fig:gap}
    (Color online) 
    Single-particle gap $\Delta$ from the insulating solutions and the 
    pseudogap from the metallic solutions vs.\ the interaction $U$.
    The determination of the pseudogap is explained in Sec.\ 
    \ref{sec:dyn:met}.
    For comparison the results from perturbation theory (PT) up to
    $\mathcal{O}(U^{-2})$ in the insulating regime are
    also shown.\cite{Eastwood-2003,endnote59}}
\end{figure}
The results are plotted in Fig.\ \ref{fig:gap} as a function
of $U$. As expected, the gap $\Delta$ decreases continuously with decreasing 
$U$ until it vanishes at a critical value $U_{\text{c1}}$. 
For $U\geq 3.4D$ our $\Delta(U)$ agrees well
with the results from perturbation theory expanding in powers of
$\nicefrac{t}{U}$.\cite{Eastwood-2003} 

Closer to the vanishing gap the behavior of $\Delta(U)$ is not linear
anymore but displays a significant curvature. Qualitatively, this
is similar to what has been observed previously in iterated perturbation theory
\cite{Rozenberg-1994} (IPT), by the local moment approach\cite{Logan-1997} 
(LMA), or by the  Lanczos approach \mbox{D-DMRG}.\cite{Garcia-2004}
This observation leads us to fit a power law
\begin{equation}
  f_{\Delta}(U)=(U-U_{\text{c1}})^{\zeta}[a_{1}+a_{2}(U-U_{\text{c1}})]
\end{equation}
with the fit parameters $a_i$, $U_{\text{c1}}$, and $\zeta$, which denotes the
leading exponent of $\Delta(U)$ close to $U_{\text{c1}}$. From this
fit $U_{\text{c1}}=(2.38\pm 0.02)D$ and $\zeta=0.72\pm 0.05$ is found.
Thus, our value for $U_{\text{c1}}$ agrees excellently with most of the
previous results, see Tab.~\ref{tab:u_critical}.

\subsubsection{Self-Energy}
\label{sssec:ins:selfenergy}

The self-energy $\Sigma$ is not required in the self-consistency cycle
shown in Fig.\ \ref{fig:sc:bethe}, but it is accessible
via Eqs.\ (\ref{eq:self-energy},\ref{eq:dyson-siam}).
By means of the self-consistently determined full propagator $G(\omega)$ and 
the bare propagator $g^0(\omega)$ of the SIAM we compute
\begin{equation}
  \label{eq:selfenergy}
  \Sigma(\omega)
  = \frac{1}{g^0(\omega)} - \frac{1}{G(\omega)}.
\end{equation}
Note, that this way to access the self-energy can be prone to numerical
problems. If the self-energy is a small quantity it can acquire a
substantial relative error if it is computed according to \eqref{eq:selfenergy}
as difference of two large quantities. In the extreme case, it can happen
that the imaginary part of the retarded self-energy computed via
\eqref{eq:selfenergy} is positive, i.e., it displays the wrong sign and
violates causality. This will be a problem in the metallic solutions
at weak coupling, see Sec.\ \ref{sec:dyn:met}.

\begin{figure}[ht]
  \centering\includegraphics[width=\columnwidth]{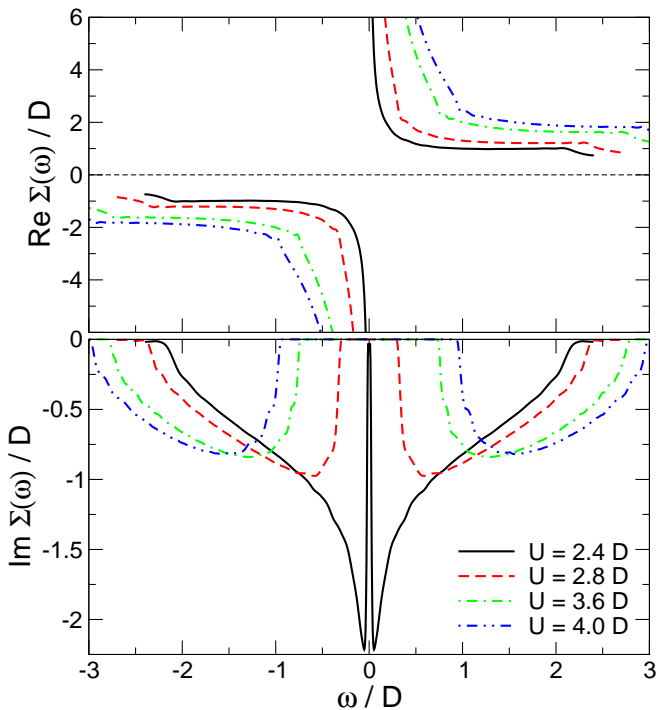}
  \caption{\label{fig:iso_selfenergy}
    (Color online) 
    Real and imaginary part of the self-energy $\Sigma(\omega)$ in the
    insulating regime. The $\delta$-peak in the imaginary part (lower panel)
    is not shown.}
\end{figure}

The indirectly calculated self-energy in the insulating regime displays the
expected features, see Fig.~\ref{fig:iso_selfenergy}. Except for a
$\delta$-peak in the imaginary part at $\omega=0$, $\Sigma(\omega)$ is strictly
real inside the gap
\begin{equation}
  \mathfrak{Im}\,\Sigma(\omega) =
  -c \pi \delta(\omega) \quad\text{for}\quad |\omega|\leq\Delta\ ,
\end{equation}
with $c>0$. Correspondingly, the real part of $\Sigma(\omega)$ shows an 
$1/\omega$ behavior:
$\mathfrak{Re}\,\Sigma(\omega) = c/\omega +\mathcal{O}(\omega)$.

\subsubsection{Momentum-Resolved Spectral Densities}
\label{sssec:ins:momres}

So far we discussed the local spectral density $\rho(\omega)$ which is
related to the imaginary part of the local propagator $G_{ii}(\omega)$.
This is the quantity which matters in the self-consistency of the
DMFT approach. But once a self-consistent solution is found one can 
go a step further towards spatial correlations. While the self-energy
and the skeleton diagrams are local in DMFT (see above) one-particle reducible
quantities like the propagators are not. Hence it makes sense to discuss
$G_{ij}$. Because we study a translationally invariant system we study
the propagator in momentum space $G_k(\omega)$.\footnote{On the Bethe
lattice there is no momentum. So in fact, we consider a translationally 
invariant  lattice with a semi-elliptic local DOS $\rho^0(\omega)$.}
 Its Dyson equation reads
\begin{equation}
\label{eq:gkomega}
G_k(\omega) = \frac{1}{\omega-\epsilon_k-\Sigma(\omega)}.
\end{equation}
By varying the bare fermionic energy $\epsilon_k$ we can assess the
effect of different momenta. In particular, if we change $\epsilon_k$
from negative to positive values we can view the change of the
single-particle response on passing from the dynamics of a hole
to the one of an electron.

\begin{figure}[t]
  \includegraphics[width=\columnwidth]{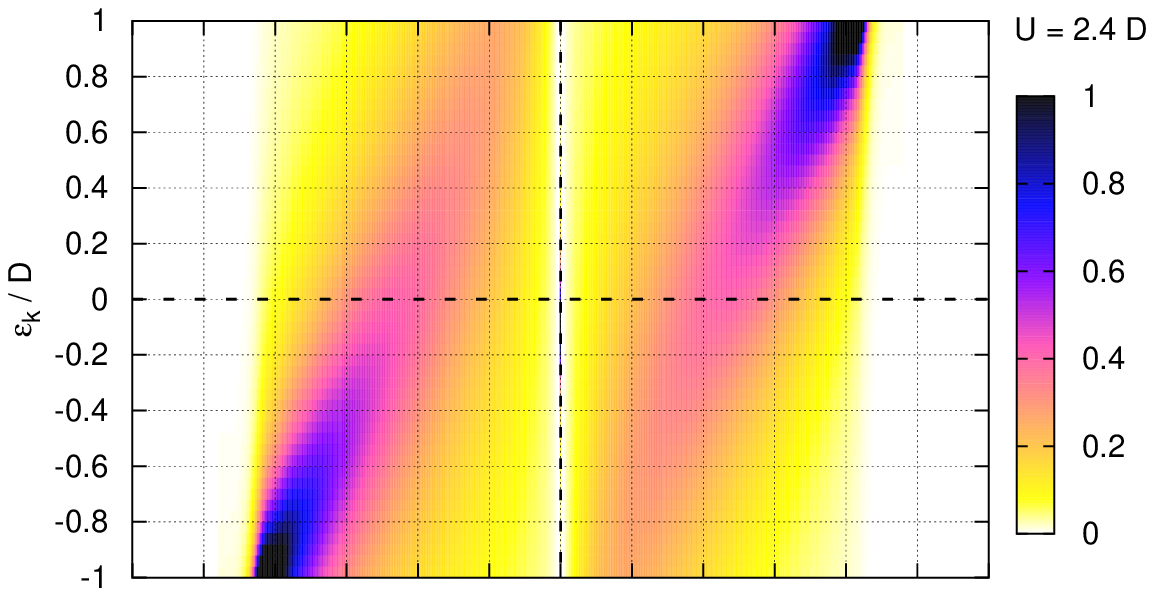}
  \includegraphics[width=\columnwidth]{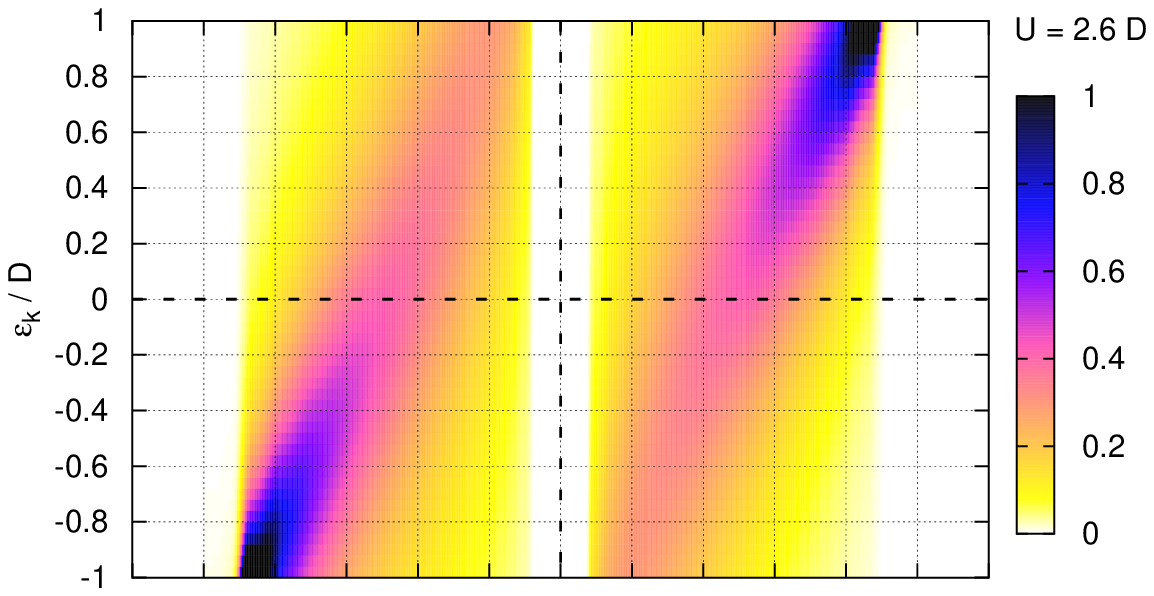}
  \includegraphics[width=\columnwidth]{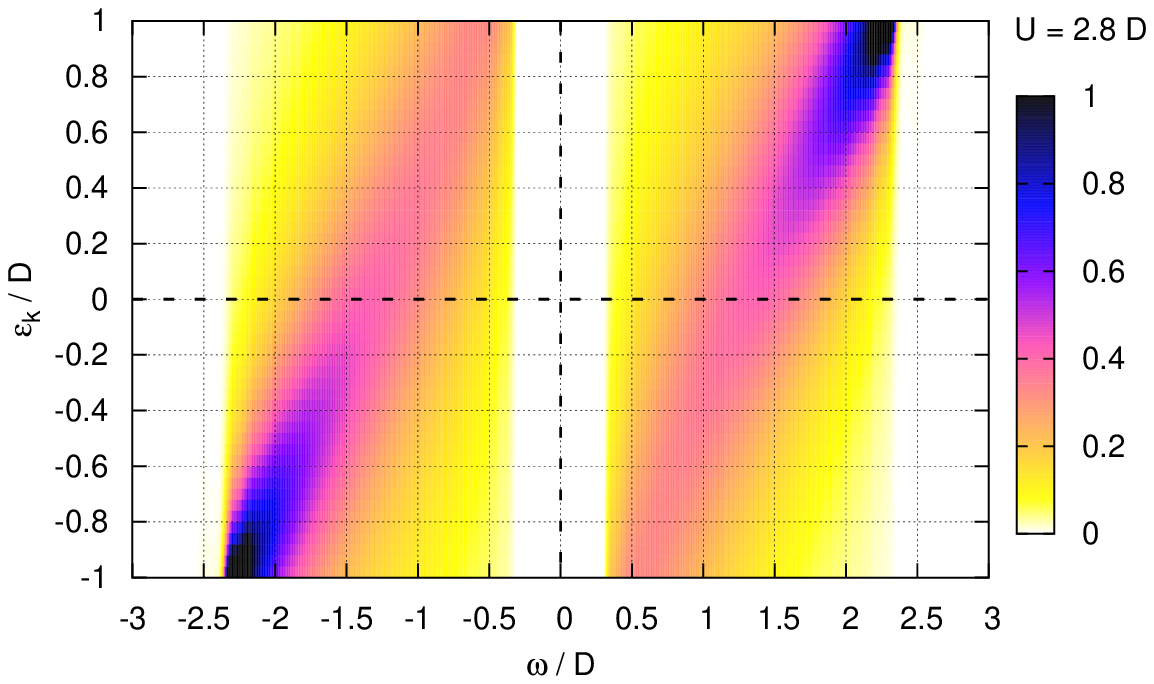}
  \caption{\label{fig:iso:1}
    (Color online)
    Spectral densities $\rho(\omega, \epsilon_k)$ in the insulating regime for
    $U=2.4D$ (top), $U=2.6D$ (middle), and $U=2.8D$ (bottom).}
\end{figure}
\begin{figure}[t]
  \includegraphics[width=\columnwidth]{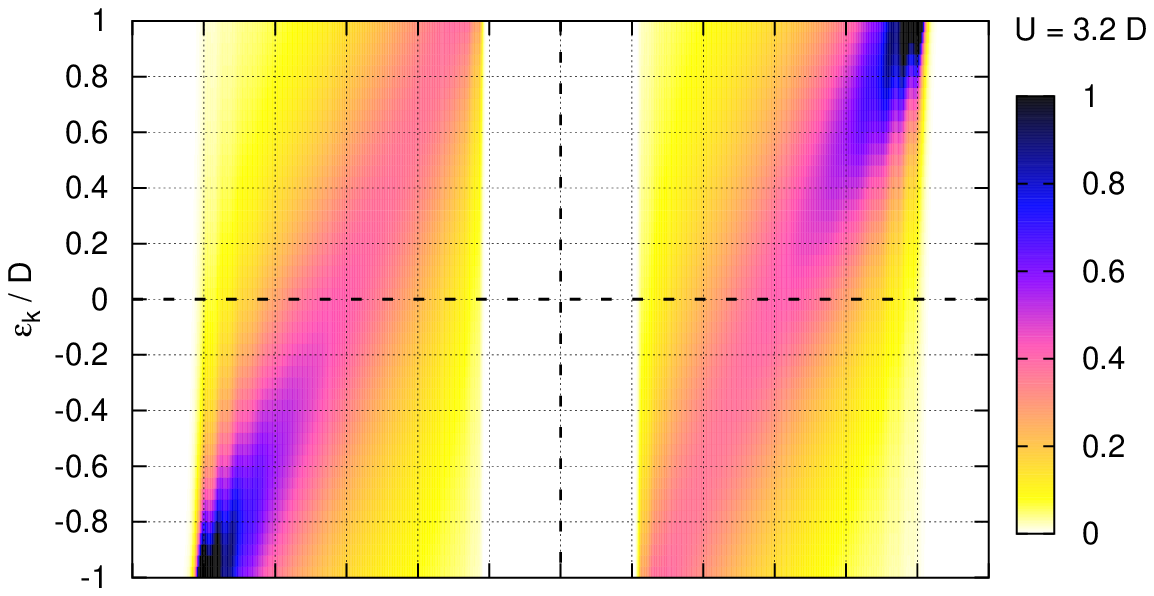}
  \includegraphics[width=\columnwidth]{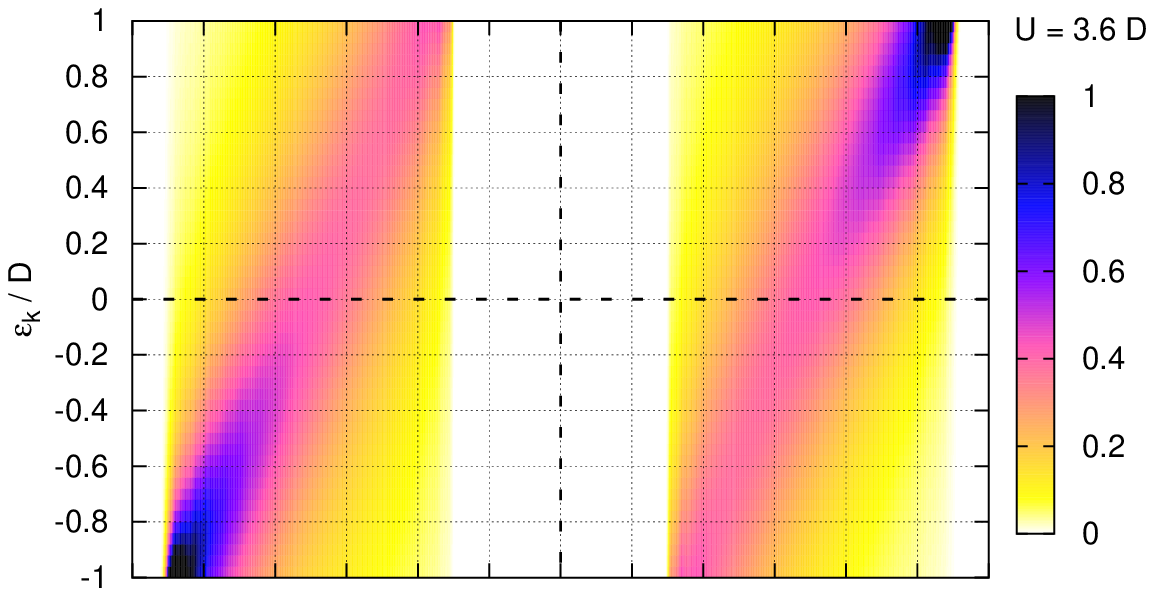}
  \includegraphics[width=\columnwidth]{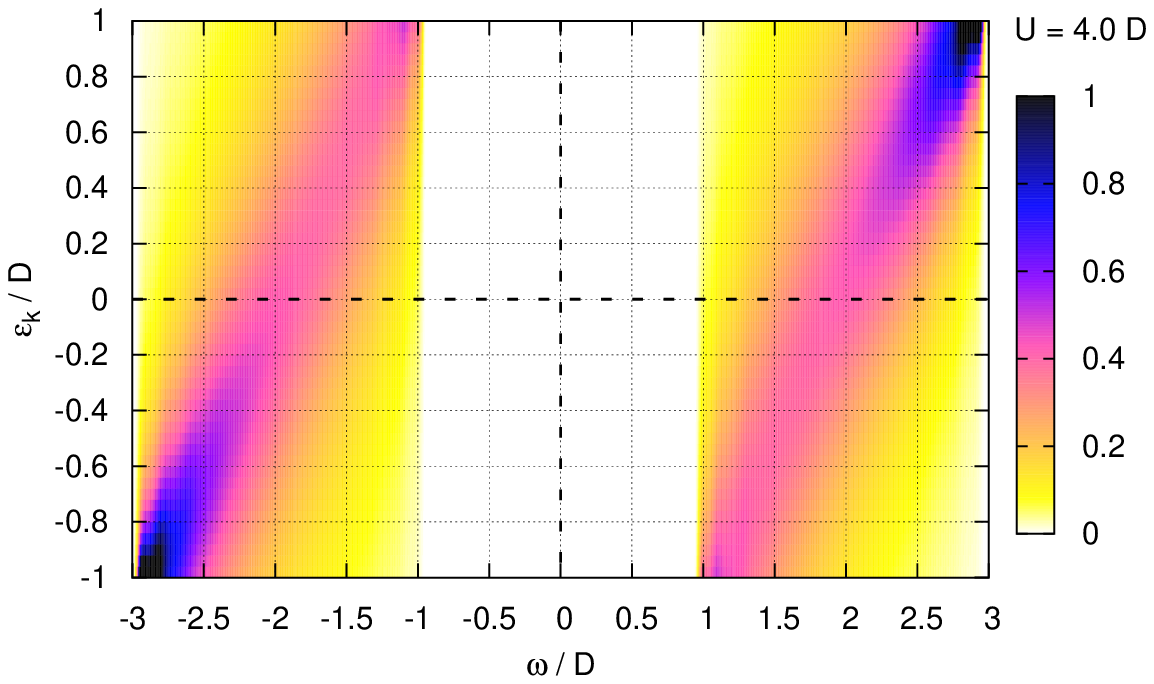}
  \caption{\label{fig:iso:2}
    (Color online)
    Spectral densities $\rho(\omega, \epsilon_k)$ in the insulating regime for
    $U=3.2D$ (top), $U=3.6D$ (middle), and $U=4D$ (bottom).}
\end{figure}

Figs.~\ref{fig:iso:1} and \ref{fig:iso:2} display the
spectral density of $G_k(\omega)$
\begin{equation}
\label{eq:rhokomega}
\rho(\omega, \epsilon_k) := - \frac{1}{\pi} \mathfrak{Im} G_k(\omega).
\end{equation}
Tiny wiggles in the intensity as function of the frequency
are numerical artifacts of the self-energy computed via \eqref{eq:selfenergy}.
The gap between the lower and the upper Hubbard band is clearly visible.
In each Hubbard band there is a ridge of high density running from
the upper right to the lower left edge. Neglecting the strong scattering
we can interpret this ridge as the dispersion of a moving double occupancy
in the upper Hubbard band. Correspondingly, there is a moving empty site
in the lower Hubbard band. Since the band width of each Hubbard band
is approximately equal to the band width of the bare dispersion  we deduce
that the hopping of the double occupancy or the empty site, respectively,
equals the bare hopping. This fact has been exploited also in the perturbative
treatment.\cite{Eastwood-2003}

\begin{figure}[ht]
  \includegraphics[width=\columnwidth]{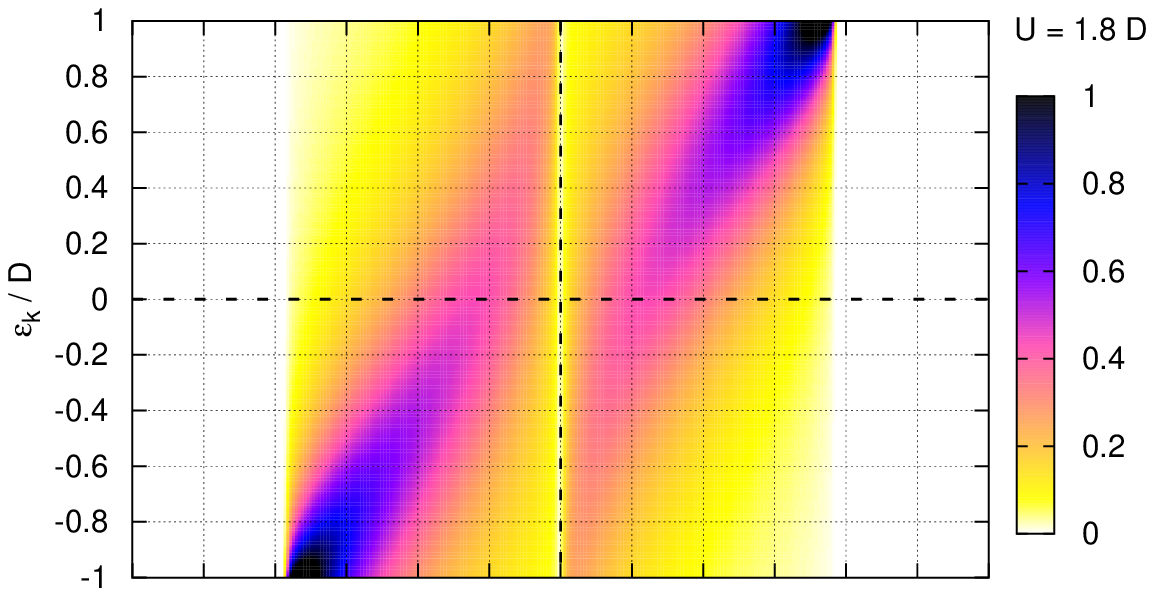}
  \includegraphics[width=\columnwidth]{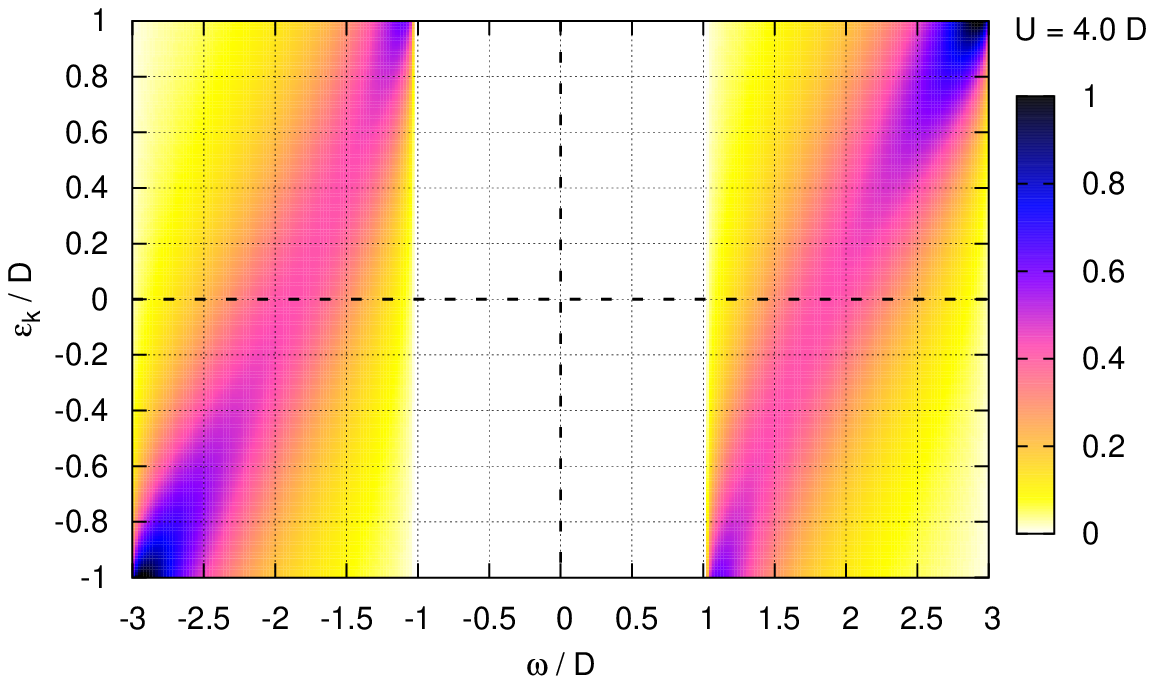}
  \caption{\label{fig:hubbard}
    (Color online)
    Spectral densities $\rho(\omega, \epsilon_k)$ in the insulating regime 
    according to Hubbard's approximation in Ref.\ \onlinecite{hubba64b} for
    $U=1.8D$ (upper panel) and $U=4.0D$ (lower panel)}
\end{figure}

Note, that the first approximation to this problem obtained by
Hubbard \cite{hubba63} yields only \emph{half} the band width for
each Hubbard band. But the famous improved solution in Hubbard's
third paper \cite{hubba64b} displays the correct band width
for strong coupling. This treatment also includes scattering
processes which take the strong scattering of the additional fermion
into account. Hence it is very instructive to compare our numerically
exact result with the approximate solution in Ref.\ \onlinecite{hubba64b}.
Fig.\ \ref{fig:hubbard} shows two examples
for an almost vanishing gap at $U=1.8D$ and for a large gap at $U=4.0D$.
The similarity to the exact results in Fig.\ \ref{fig:iso:1}, upper panel,
and in Fig.\ \ref{fig:iso:2}, lower panel, is striking. Both results,
exact and approximate, show strongly scattered, hence broadened, excitations
in the Hubbard bands. The ridges behave very similarly in shape and height.
In particular for the large gap the results agree well.

For the smaller value of the insulating gap, the most important
difference is that the value of $U$ differs. But if we accept that
the interaction value in Hubbard's approximation needs to be renormalized
in a certain way the spectral densities behave qualitatively very similar.
It must be stressed, however, that the results are quantitatively
different. Comparing the upper left panel in Fig.\ \ref{fig:iso_dens}
with the corresponding result in Ref.\ \onlinecite{hubba64b}
one notes that the spectral weight in the exact solution is found
further away from the Fermi level at $\omega=0$ while in Hubbard's
solution it is shifted towards $\omega=0$.


\subsection{Metal}
\label{sec:dyn:met}

\subsubsection{Local Spectral Densities}
\label{sssec:met:lsd}

\begin{figure}[ht]
  \centering\includegraphics[width=\columnwidth]{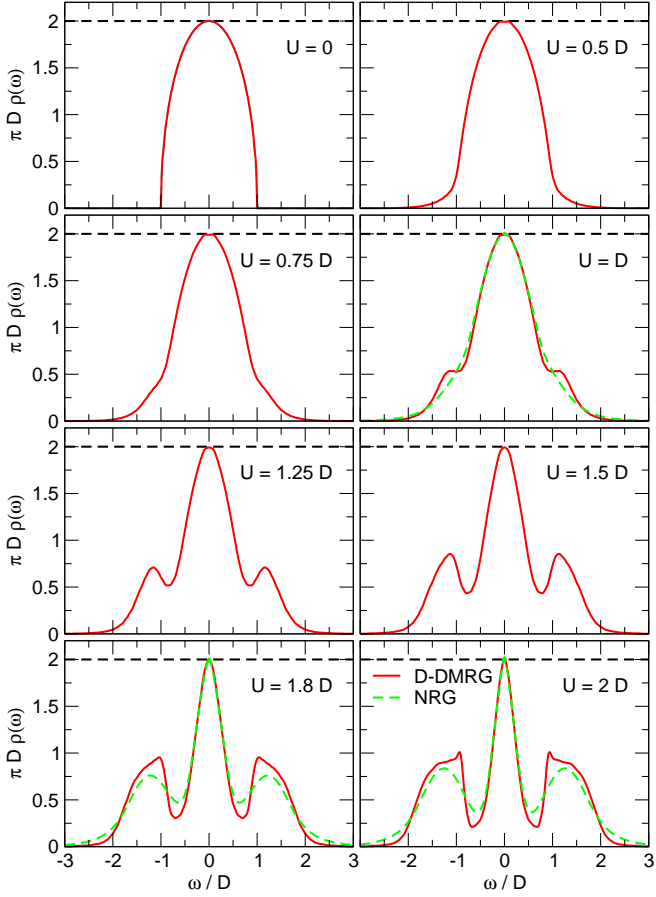}
  \caption{\label{fig:met_dens_far}
    (Color online)
    Spectral densities $\rho(\omega)$
    in the metallic regime calculated with a chain of 160 fermionic sites.
    The dashed lines show results of a DMFT approach based on 
    the numerical renormalization group (NRG) as impurity 
    solver.\cite{Bulla-2004}}
\end{figure}
\begin{figure}[ht]
  \centering\includegraphics[width=\columnwidth]{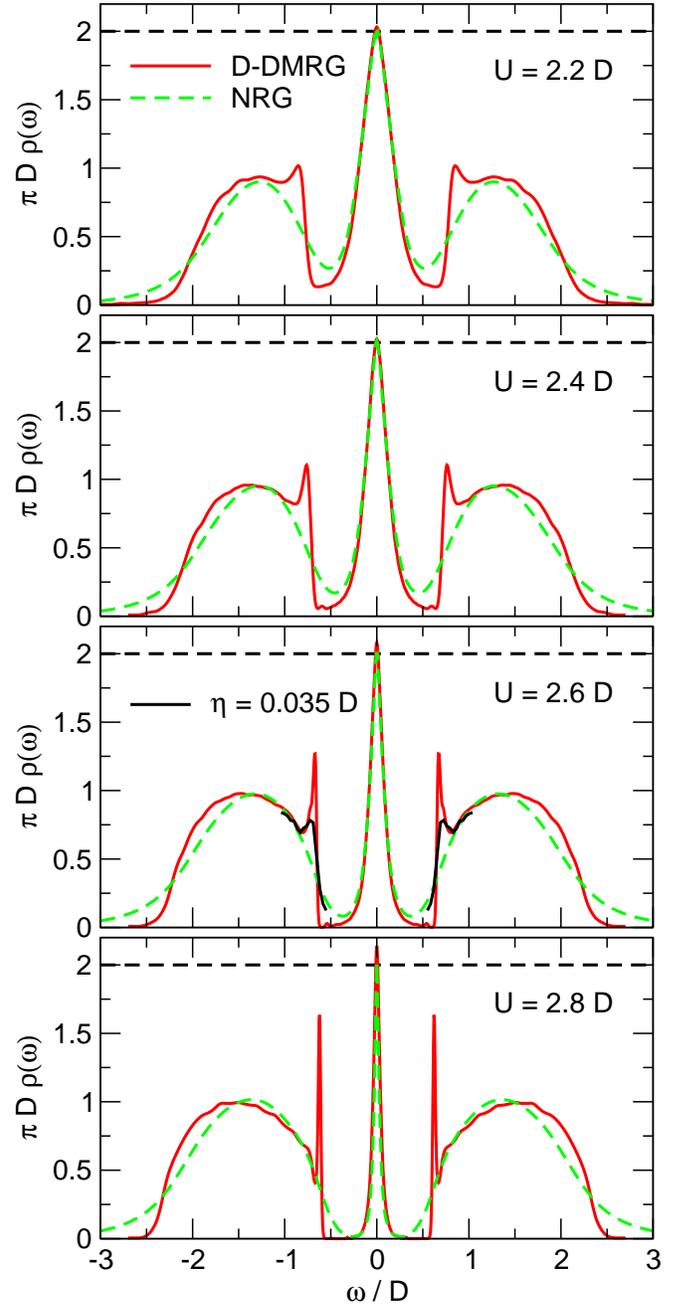}
  \caption{\label{fig:met_dens_close}
    (Color online) 
    Spectral densities $\rho(\omega)$
    in the metallic regime close to the metal-to-insulator transition.
     The dashed lines show
    results of an NRG-based DMFT approach.\cite{Bulla-2004}
    For $U=2.6D$, the dark solid lines show the raw \mbox{D-DMRG} data
    broadened with $\eta=0.035D$ around the inner Hubbard band
    edges, for discussion see Sec.\ \ref{sssec:met:sisp}}
\end{figure}
Figure~\ref{fig:met_dens_far} shows the self-consistently calculated spectral
densities $\rho(\omega)$ well in the metallic regime ($0\leq U\leq 2D$). 
Figure~\ref{fig:met_dens_close} shows the corresponding
densities $\rho(\omega)$ closer to the transition to the
insulating solution at $U=U_\text{c2}\approx 3.0D$. 
We used $160$
fermionic sites (including one site representing the impurity) in the
\mbox{D-DMRG} impurity solver. This implies a tiny bias towards
the metallic solution, see Sec.\ \ref{sssc:numer-stabl}.
The raw data is obtained on an equidistant frequency grid
with step size $\Delta\omega_i=0.1D$ using a constant broadening
$\eta=0.1D$.  It is deconvolved using the LB algorithm.\cite{Raas-2005a} 
The spectral densities in 
Figs.~\ref{fig:met_dens_far}~and~\ref{fig:met_dens_close} qualitatively display
many of the features known from previous analytical
\cite{Eastwood-2003,Gebhard-2003} and numerical
\cite{Georges-1992,Jarrell-1992,Rozenberg-1992,Georges-1993,Caffarel-1994,Sakai-1994,Bulla-1999,Bulla-2001,Garcia-2004} investigations.

Starting from the semi-elliptic DOS \eqref{eq:bethe_dos} at $U=0$
the central peak becomes narrower and narrower on increasing $U$.
The diminishing width is the Kondo energy scale $T_\text{K}$. The
weight in the central peak is the quasiparticle weight $Z$ which
we will discuss below.

The height of the central peak does not change. The quasiparticle peak is
pinned at the Fermi energy to its non-interacting value
\begin{equation}
  \label{eq:pin}
  \rho(\omega=0) = \rho^0(\omega=0) = \frac{2}{\pi D}
\end{equation}
as required by Luttinger's theorem for a momentum independent
self-energy.\cite{Mueller-Hartmann-1989b}
As the DOS at zero frequency is not pre-determined in our approach,
the pinning criterion \eqref{eq:pin} provides a sensitive check for the 
accuracy of the numerical solutions in the metallic regime.
This criterion is very well satisfied, see Fig.\ 
\ref{fig:met_dens_far} and the upper panels in Fig.\ 
\ref{fig:met_dens_close}. 
A maximum relative error of $3\%$ occurs for interaction values close
to the transition to the insulator, see lower panels in Fig.\ 
\ref{fig:met_dens_close}. 
The fulfillment of the pinning is an important prerequisite for an accurate
determination of the quasiparticle weight $Z$ below.

Beyond $U\approx D$, the DOS shows side features.
They are just shoulders which develop into the lower
and upper Hubbard bands at larger interaction. 
As expected, they appear roughly at 
$\omega=\pm\nicefrac{U}{2}$.  At $U\approx 2D$ the Hubbard bands are well
separated from the quasiparticle peak by a precursor of the gap $\Delta$ in 
the insulator: a pseudogap is formed.
This pseudogap is determined from $U=2.4D$ onwards by background
fits as shown in Fig.\ \ref{fig:sidepeak_weight_fit}.
The resulting values and their uncertainties are included in 
Fig.\ \ref{fig:gap}.
Their extrapolation corroborates that the pseudogap evolves
continuously to the insulating gap at $U_\text{c2}\approx 3D$
where the metallic solution becomes unstable towards the insulating
solution. 

The well pronounced pseudogap in the metallic solutions close to the transition
reveals clearly that the quasiparticle peak becomes isolated from the
Hubbard bands at higher energies. We demonstrated in Fig.\ 2 in
Ref.\ \onlinecite{Karski-2005} that the spectral densities of the metallic 
and of the co-existing insulating solution almost coincide at higher energies.
This means that except for the quasiparticle peak and the inner edges
of the Hubbard bands the insulating and the metallic solutions 
are almost identical. The two phases differ in their behavior at
small and intermediate energies. This phenomenon has been
baptized \emph{separation of energy scales}. The assumption of its validity
is at the basis of the numerical projective self-consistent method 
\cite{Moeller-1995,Georges-1996} and of 
Kotliar's analytical Landau theory.\cite{kotli99}
Hence, the fact that our all-numerical treatment indeed displays
the separation of energy scales represents a valuable confirmation
of this important hypothesis.

It is very instructive to compare our findings for $\rho(\omega)$
with the NRG data from Ref.~\onlinecite{Bulla-2004}. The quantitative
comparison is depicted in some of the panels in Fig.\ 
\ref{fig:met_dens_far} and in Fig.\  \ref{fig:met_dens_close}.
The agreement is very good in the close vicinity of the quasiparticle peak,
i.e., at low energies. But significant deviations are discernible for
intermediate and higher frequencies. The Hubbard bands obtained by the
\mbox{D-DMRG} are sharper compared to the NRG results. Hence their
precursors become visible already at lower values of $U$, see
Fig.\ \ref{fig:met_dens_far}. Furthermore, the Hubbard bands in \mbox{D-DMRG}
do not have significant tails at higher energies. 
This difference stems from the broadening
proportional to the frequency which is inherent to the NRG
algorithm.\cite{sakai89,costi90,Bulla-1999,Bulla-2001,Raas-2004} 
Similar differences were also reported in
Ref.~\onlinecite{Gebhard-2003} where perturbative results are compared
with NRG data.

\begin{figure}[ht]
  \centering\includegraphics[width=\columnwidth]{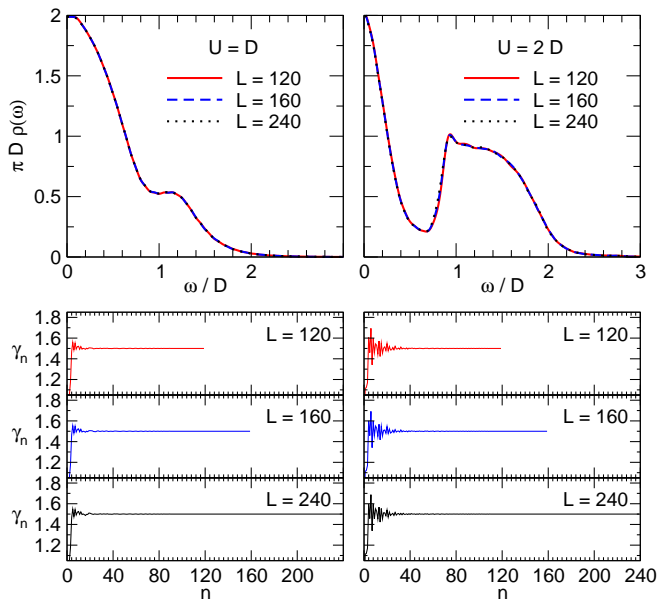}
  \caption{\label{fig:fsites_100_200}
    (Color online) 
    Part of the spectral density $\rho(\omega)$
    and the associated bath parameters $\gamma_n$ calculated
    self-consistently for various numbers $L$ of fermionic sites.
    \emph{Left panel:}  $U=D$.
    \emph{Right panel:} $U=2D$.}
\end{figure}

In order to investigate a possible influence of finite-size effects, we 
consider the self-consistent spectral densities and the associated bath
parameters $\gamma_n$, cf.~Eq.~\eqref{eq:ham_siam}, for various sizes
of the bath.
Fig.~\ref{fig:fsites_100_200} shows the results for $L=120$, $160$ and $240$
fermionic sites well in the metallic regime for $U=D$ and $U=2D$. If the finite
broadening of the \mbox{D-DMRG} raw data is chosen according to
Eq.~\eqref{eq:fsites_broadening} the spectral densities deconvolved by LB
are not affected significantly by the number of bath sites. We observe neither
qualitative nor significant quantitative differences. This observation applies
also to the self-consistently determined bath parameters $\gamma_n$. Above
$n\approx80$ almost all bath parameters take the same value which
determines the band width, see App.\ \ref{app:cf}. 

We conclude that a system size
of $80$ fermionic sites is already sufficient to capture quantitatively the
features of the spectral properties well in the metallic regime.
Nevertheless, a small improvement in the pinning criterion
can be observed if more fermionic sites are considered, see
Ref.~\onlinecite{Karski-2004}). This improvement arises from a better
approximation of the thermodynamic limit which makes the deconvolution using 
the LB algorithm more robust. But the system size cannot be chosen too
large in practice because the \mbox{D-DMRG} calculation will acquire systematic
errors for large systems and given computational resources.
Our D-DMRG calculations were all done with 128 states, partly checked by
runs with 256 states. But we have not noticed significant deviations
between the runs with 128 and those with 256 states.

\subsubsection{Sharp Inner Side Peaks}
\label{sssec:met:sisp}

For strong interactions ($U>2D$) close to the metal-to-insulator transition the
metallic solutions show sharp peaks at the inner edges of the Hubbard bands, 
see Fig.~\ref{fig:met_dens_far} for $U=2D$ and Fig.~\ref{fig:met_dens_close}. 
These peaks become sharper and sharper on increasing the interaction $U$.
Note, that their height cannot (and does not) exceed the maximum value
of the non-interacting DOS because the interacting DOS is determined
 from the Dyson equation \eqref{eq:dyson-lattice}. 
It implies that $\rho(\omega)$ results
from $\rho^0(\omega)$ by a shift in frequency due to $\mathfrak{Re}\Sigma$
and by a Lorentzian broadening due to $\mathfrak{Im}\Sigma$ so that
$\rho(\omega)$ is bounded by the maximum of $\rho^0(\omega)$.
The sharp features do not represent diverging singularities.

In order to properly resolve the sharp side peaks as well as
the narrow quasiparticle peak, we choose $\Delta\omega_i=\eta_i=0.05D$ and
partially $\Delta\omega_i=\eta_i=0.025D$ in the relevant frequency regions.
Furthermore, we place the computed points of the raw data at the frequency
of the maximum of the side peaks. 

\begin{figure}[ht]
  \centering\includegraphics[width=\columnwidth]{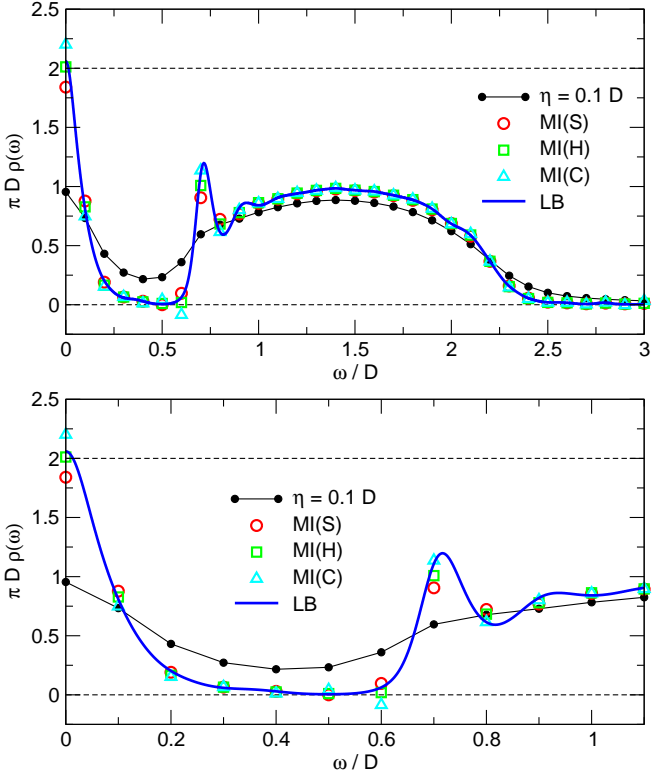}
  \caption{\label{fig:deconv}
    (Color online)
    Comparison of various deconvolution schemes for $U=2.6D$. 
    The solid line with filled circles depicts the raw data.
    MI refers to linear matrix
    inversion for which we assume that the DOS consists of a set of 
    $\delta$-functions [spikes, MI(S): circles] or that it is piecewise 
    constant [histogram, MI(H): squares] or that it is
    piecewise linear and continuous [MI(C): triangles], for details see
    Ref.~\onlinecite{Raas-2005a}.
    LB stands for the continuous positive result obtained by the non-linear
    LB deconvolution.
    \emph{Upper panel:} complete frequency interval of the symmetric DOS.
    \emph{Lower panel:} details at lower frequencies.}
\end{figure}
In Fig.\ \ref{fig:deconv} we show that the sharp peaks at the inner
band edges are not an artifact of the deconvolution algorithm since
a large variety of deconvolution schemes  essentially provides the same
peak. Furthermore, the side peaks are already discernible in the 
raw data at small broadening (dark solid lines) 
shown in Fig.~\ref{fig:met_dens_close} for $U=2.6D$.

We checked that different initial hybridization functions led to the
same self-consistent solutions after sufficient iterations
so that we exclude that the side peaks
are artifacts of the self-consistency procedure. Moreover, Nishimoto
et al.\ found very similar side peaks by a completely independent
implementation of DMFT by D-DMRG.\cite{nishi06}
Hence we conclude that the sharp side peaks resolved in our \mbox{D-DMRG}
approach are part and parcel of the metallic spectral density of the
half-filled Hubbard model close to the metal-to-insulator transition.

\begin{figure}[ht]
  \centering\includegraphics[width=\columnwidth]{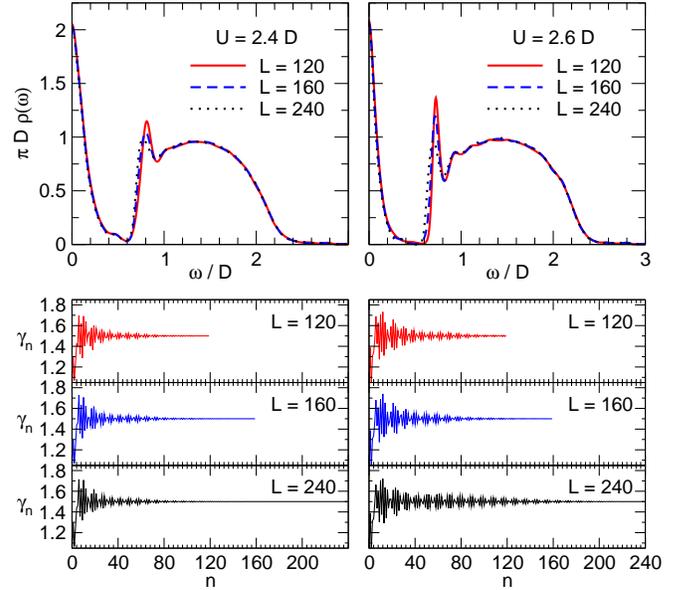}
  \caption{\label{fig:fsites_240_260}
    (Color online) 
    Part of the spectral density $\rho(\omega)$
    and the associated bath parameters $\gamma_n$ calculated
    self-consistently for various numbers $L$ of fermionic sites.
    \emph{Left:}  $U=2.4D$.
    \emph{Right:} $U=2.6D$.}
\end{figure}
We repeated the investigation of the possible influence of the system size 
in Fig.\ \ref{fig:fsites_100_200} close to the metal-to-insulator transition. 
In Fig.~\ref{fig:fsites_240_260} we compare spectral densities for $U=2.4D$ and
$U=2.6D$ and their bath parameters $\gamma_n$
determined self-consistently for $120$, $160$, and $240$ fermionic sites
with $\Delta\omega_i=\eta_i=0.1D$. Smaller values for $\Delta\omega$ and
$\eta$ turned out not to be feasible for the large system sizes
due to limited computational resources. 
Thus, we accept a slight loss in the resolution of the side peaks 
compared to the spectral densities shown in Fig.~\ref{fig:met_dens_close} 
where calculations with $\Delta\omega_i=\eta_i=0.025D$ were performed
for frequencies around the side peak positions.

We see in Fig.~\ref{fig:fsites_240_260} that the precise position 
and the shape of the side peaks depend slightly on the size of 
the bath considered in the numerical calculations. The other, less
sharp features, like the central peak and the Hubbard band are very robust
against the changes of the bath size. It is very instructive to
look at the bath paramters $\gamma_n$. For  $U=2.4D$ they do not change
on passing from $L=120$ over  $L=160$ to  $L=240$. The oscillations
visible from smaller values of $n\lessapprox 80$ are essentially the same
for the three system sizes. They do not extend to larger $n$ for
larger $L$. For $U=2.6D$, however, the bath parameter change
for larger systems. The oscillations extend to higher $n$ for $L=240$ 
than for lower $L$. For $L=240$ they appear to have converged for large
$n$ so that we assume that increasing $L$ further would only extend the
region of almost constant $\gamma_n$.

We conclude, that sufficiently large baths are essential for a quantitative
description of the spectral densities and of the sharp features at the
inner edges of the Hubbard bands in particular.
This is especially important close to the metal-to-insulator transition. 
Our results confirm the existence of the sharp side peaks
and provide their accurate determination
for all $U\leq 2.2D$. For $U>2.2D$ the possibility of small inaccuracies of 
the order of the deviations shown in the upper panels of 
Fig.~\ref{fig:fsites_240_260} must be kept in mind.

The only previous evidence for sharp side peaks in numerical investigations
were weak shoulders in NRG\cite{Bulla-1999} and quantum Monte Carlo (QMC)\cite{Bluemer-2002} data.
The inherent broadening in the NRG calculations and the necessity to continue 
the QMC data from imaginary times to real frequencies have both a tendency to 
smear out sharp features away from the Fermi level, cf.\ Ref.\ 
\onlinecite{Raas-2004}. Hence we tend to interprete the shoulders in
Refs.~\onlinecite{Bulla-1999}~and~\onlinecite{Bluemer-2002}
as broadened remnants of the sharp peaks resolved in the present \mbox{D-DMRG}
calculation.

Besides the numerical approaches semi-analytical approaches, 
the iterated perturbation theory (IPT)\cite{Georges-1996} and the non-crossing
approximation (NCA),\cite{Pruschke-1993} 
display similar behavior of the spectral densities 
near the inner Hubbard band edges. The IPT results show a substantial 
accumulation of  spectral  weight at the inner band edges which differs very 
much from our findings: the weight is much larger and the features are by far 
not as sharp as those presented here. They carry even more spectral weight than
the quasiparticle peak and they are accompanied by spike structures over the 
entire Hubbard bands in contrast again to our findings. In view of these
differences and because IPT is an approximate diagrammatic approach we conclude
that the IPT features do not correspond to the ones presented here.

The NCA data in Ref.~\onlinecite{Pruschke-1993} is qualitatively much
closer to our data as far as the side peaks are concerned. 
Since the NCA violates the Fermi liquid properties for temperatures lower than 
the Kondo scale \cite{Mueller-Hartmann-1984} reliable conclusions 
can only be drawn for sufficiently large temperatures. Moreover, the peaks
are found in the insulating phase so that it is an open question whether 
the NCA findings are related to the D-DMRG side peaks at zero
temperature in the metallic regime.

\begin{figure}[ht]
  \centering\includegraphics[width=\columnwidth]{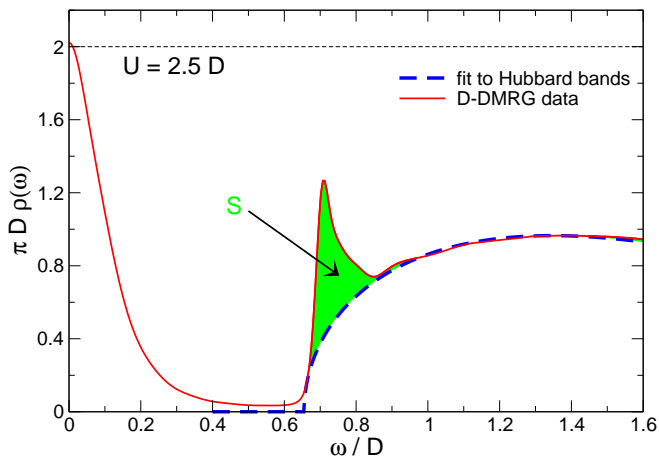}
  \caption{\label{fig:sidepeak_weight_fit}
    (Color online)
    Determination of the side peak weight $S$ illustrated for $U=2.5D$.}
\end{figure}

In order to pave the way to a physical interpretation of the
side peaks we determine its weight $S$. Motivated by the results
for insulating $\rho(\omega)$ in Sec.\ \ref{sssec:ins:lsd}
we fit an approximate square-root onset multiplied
by a quadratic polynomial to the Hubbard band. The excess weight is
identified with the spectral weight $S$ of the sharp side peak. 
This procedure is illustrated in Fig.~\ref{fig:sidepeak_weight_fit}. 

\begin{figure}[ht]
  \centering\includegraphics[width=\columnwidth]{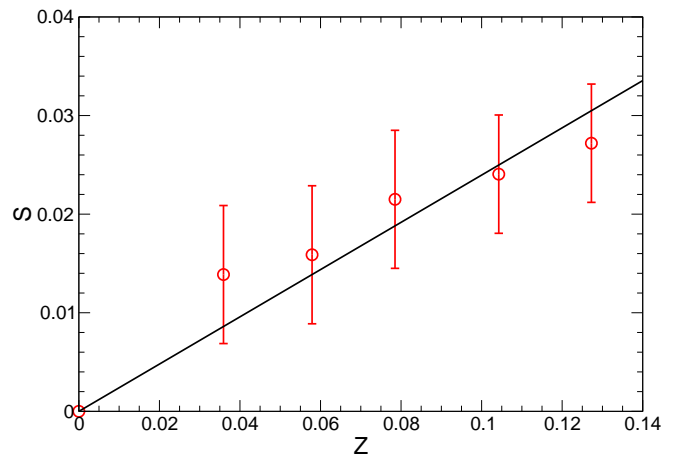}
  \caption{\label{fig:sidepeak_weight}
    Weight $S$ of the peaks at inner Hubbard band edges as a function of the
    quasiparticle weight $Z$.}
\end{figure}
Inspection of Fig.~\ref{fig:met_dens_close} suggests that $S$ vanishes
for $U\to U_\text{c2}$. This conclusion is supported by the fact that
the insulating solutions do not display a similar feature, see Sec.\
\ref{sssec:ins:lsd}. Hence  we presume that the side peak involves
quasiparticles from the central peak. The weight in this central peak
is given by $Z$ which vanishes linearly for $U\to U_\text{c2}$, see 
Sec.\ \ref{sssec:met:qpw}. So we plot the results for $S$ as function of $Z$
in Fig.\ \ref{fig:sidepeak_weight} including a point $S=0$ for $Z=0$.
The error bars result from the numerical
difficulty to resolve this sharp feature and from the analytical difficulty to
separate it from the background of the Hubbard bands.

In view of the errors the interpretation of Fig.\ \ref{fig:sidepeak_weight}
is daring. But our guiding hypothesis is that the 
surprisingly sharp side peak is the signature of a composite excitation
involving several fermionic excitations.\cite{Karski-2005}
A single quasiparticle alone cannot be at the origin of a sharp feature
because it has  a finite dispersion and a finite life-time so that
it engenders only broad features in local quantities away from the Fermi level.

One possible assumption is that the composite excitation comprises three 
quasiparticles (Note, that the total statistics must be fermionic.). But they
need to be bound or anti-bound to account for the very sharp structure
which tells us that the composite entity must be almost immobile.
Otherwise it would give rise only to a broad structure in local
spectral densities, which correspond to sums over the entire Brillouin zone.
A plausible second scenario is that a single quasiparticle is combined
with a collective mode, for instance a particle-hole pair as it occurs
when a spin is flipped. 

With these ideas in mind, we want to use Fig.\ \ref{fig:sidepeak_weight} only
to decide whether one, two or three quasiparticles are involved.
The quasiparticle weight $Z$ measures to which extent a local fermionic 
creation operator
generates a true quasiparticle including its polarization dressing.
So we presume that the contribution of a single quasiparticle gives rise
to a linear behavior $S\propto Z$ while two quasiparticles lead to 
$S\propto Z^2$, and three quasiparticles imply $S\propto Z^3$.
Hence we find in the linear fit in Fig.\ \ref{fig:sidepeak_weight} 
a hint that only one quasiparticle contributes to the side peaks.

Additionally, we presume that an immobile collective mode is involved.
Note, that due to the scaling \eqref{eq:prop-scale} in DMFT
collective modes are as such immobile because their motion requires
at least \emph{two}  single-particle propagators.
The nature, however, of the collective mode inducing the composite
feature is still an open issue which we address in ongoing investigations.

\subsubsection{Self-Energy}
\label{sssec:met:selfenergy}

\begin{figure}[ht]
  \centering\includegraphics[width=\columnwidth]{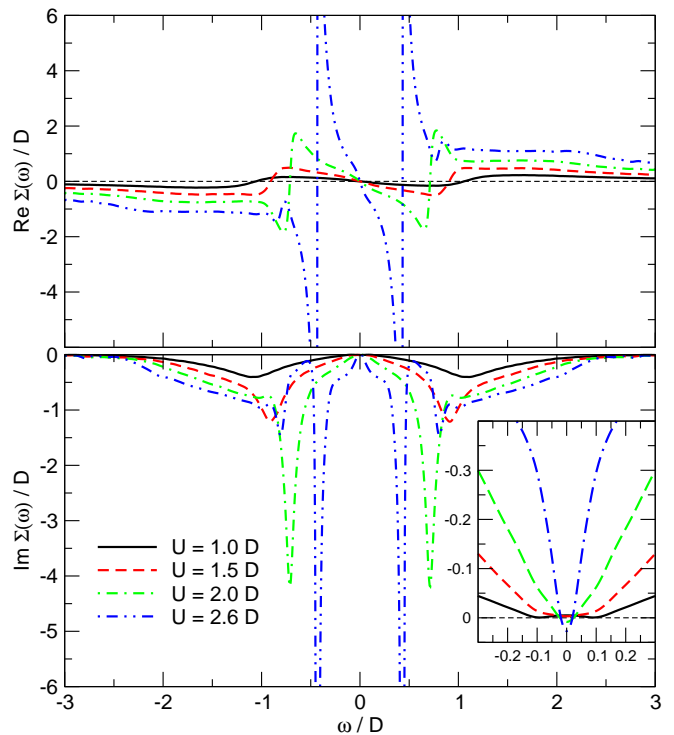}
  \caption{\label{fig:met_selfenergy}
    (Color online) 
    Real and imaginary part of the self-energy $\Sigma(\omega)$ in the metallic
    regime.
    \emph{Inset:} $\mathfrak{Im}\,\Sigma(\omega)$ on a larger scale.}
\end{figure}

In the metallic phase we use again Eq.~\eqref{eq:selfenergy}
to obtain the self-energy. The results are plotted in 
Fig.~\ref{fig:met_selfenergy}. The real and imaginary part of the
self-energy display a typical low-frequency behavior of a local Fermi liquid:
The real part shows a linear behavior with a negative slope which steepens as
$U$ increases. In this region, the imaginary part of $\Sigma(\omega)$ shows a
quadratic behavior, $\mathfrak{Im}\,\Sigma(\omega)\propto-\omega^{2}$, as
expected.\cite{Mueller-Hartmann-1989b} 
The quadratic behavior of $\mathfrak{Im}\Sigma$ is not perfectly retrieved, 
see inset of  Fig.~\ref{fig:met_selfenergy}. For low values of $U$ it seems 
that $\mathfrak{Im}\Sigma$ is proportional to a higher power than $\omega^2$
for $\omega\to0$. For large interaction values, for instance $U=2.6D$, 
the computation of the self-energy via \eqref{eq:selfenergy} even leads
to negative spectral densities (positive $\mathfrak{Im}\Sigma$
for the retarded self-energy) violating causality. This is a clear
indication for the limits of the accuracy of the present calculation.
But we stress that these difficulties are minor and do not affect the
self-consistency because they occur only when the self-energy is computed
from the full propagator $G$.

The two-peak structure in the imaginary part of $\Sigma(\omega)$ is closely
related to the three-peak structure of the spectral density given by
lower Hubbard band, central Kondo peak, and the upper Hubbard band. 
According to Eq.~\eqref{eq:selfenergy} the self-energy has a peak if the real 
and the imaginary part of $G(\omega)$ are small. For the metallic solutions, 
this is the case in the frequency regime between the Hubbard bands and the
quasiparticle peak, i.\,e.~within the pseudogap.

This relation has been previously analyzed.\cite{zhang93,kehre98b,Bulla-1999}
It turns out that the strong peaks in the self-energy are located at
position $\omega_\Sigma \propto \pm \sqrt{Z}D$, where $Z$ is the quasiparticle
weight, see next subsection.  Our data agrees very well with the
analytical arguments; the factor of proportionality is about $\sqrt{2}$.
We stress that $\omega_\Sigma \to 0$ on approaching the
metal-to-insulator transition whereas
the sharp inner side peaks remain at the inner band edges.
Hence the two peaks in the imaginary part of the 
self-energy are not directly related to the sharp side peaks.
It should be pointed out that the two-peak structure of the self-energy
contradicts the findings in Ref.\ \onlinecite{kehre98b}. 
We presume that the analytical argument misses some effect which requires
further research.

\subsubsection{Quasiparticle Weight}
\label{sssec:met:qpw}

The quasiparticle weight $Z$ is the characteristic quantity of the metallic
phase. Crudely speaking $Z$ quantifies how much of the spectral weight
is found in the central quasiparticle peak, 
cf.\ Fig.\ \ref{fig:met_dens_close} and \ref{fig:met_dens_far}.
In other words, $Z$ is the matrix element between the fermionic creation
operator applied to the ground-state and the dressed quasiparticle
excitation including its polarization cloud.
For local self-energies \cite{Mueller-Hartmann-1989b} it is given by
\begin{equation}
  Z^{-1} = 1 - 
  \left.\partial_{\omega} \Sigma(\omega)\right|_{\omega=0}.
\end{equation}
Because the self-energy is not directly accessible in our approach
 we prefer to calculate $Z$ using the derivative of the Dyson equation 
\eqref{eq:dyson-lattice} implying 
\begin{equation}\label{eq:z_green}
  Z^{-1} = \frac{D^{2}}{2} \left.\partial_{\omega}G(\omega)\right|_{\omega=0},
\end{equation}
where $\partial_{\omega}G^0(\omega)|_{\omega=0}=\nicefrac{2}{D^2}$ 
has been used.
By Eq.~\eqref{eq:z_green} we avoid the numerical errors
occurring if the self-energy is determined via \eqref{eq:selfenergy},
 see Sec.\ \ref{sssec:met:selfenergy}.

The derivative $\partial_{\omega}G(\omega)|_{\omega=0}$ in 
Eq.~\eqref{eq:z_green} is real because the imaginary part displays an
extremum at $\omega=0$.
The derivative of $\mathfrak{Re}\,G(\omega)$ is obtained reliably from the
coefficient $a$ of the fit function
\begin{equation}
  f_{\mathfrak{Re}\,G}(\omega) = a\omega + b\omega^3
\end{equation}
where the absence of even orders due to  particle-hole symmetry has been
taken into account. The fit is done in the intervals around the origin
which decrease for $U\to U_\text{c2}$ roughly linearly; for $U=2D$ it
is $[-0.1D,0.1D]$. 
The derivative cannot be determined robustly as ratio of differences because
$\mathfrak{Re}\,G(\omega)$ displays weak oscillations resulting
via the  Kramers-Kronig transformation
from weak wiggles in $\mathfrak{Im}\,G(\omega)$ due to the 
deconvolution procedure, see for instance the lowest panel
in Fig.\ \ref{fig:met_dens_close} around $\omega=D$.

\begin{figure}[t]
  \centering\includegraphics[width=\columnwidth]{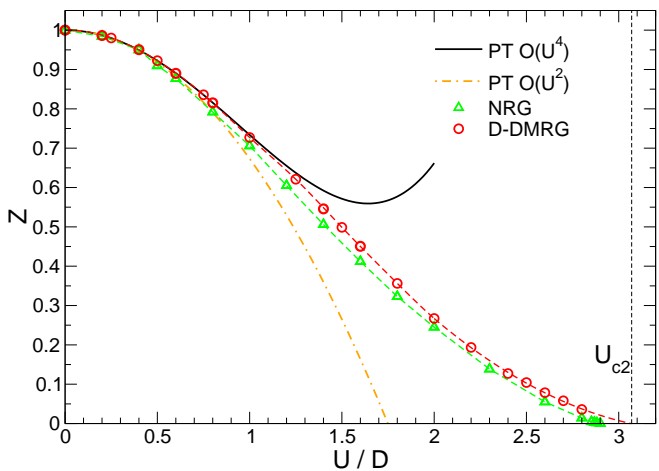}
  \caption{\label{fig:pt_weight}
    (Color online) 
    Quasiparticle weight $Z$ as a function of $U$
    compared to perturbative results\cite{Gebhard-2003} (PT) up to
    $\mathcal{O}(U^{4})$ and to  NRG data.\cite{Bulla-1999}}
\end{figure}

In Fig.~\ref{fig:pt_weight} the quasiparticle weight $Z$ is depicted as a 
function of $U$. Starting from $Z=1$ at $U=0$, $Z$ decays as $U$ increases.
Our results agree excellently with perturbation theory where it is 
applicable $U\lessapprox D$. The NRG data is lower for all values
of $U$. We deduce that the broadening of the spectral weights proportional
to $\omega$, inherent to NRG, has a non-negligible influence on 
the quasiparticle weight. Qualitatively, however, the behavior
of $Z(U)$ is very similar.

From the trend of $Z(U)$ one expects that $Z$ vanishes continuously at a finite
critical value $U_{\text{c2}}$. We cannot approach this value
arbitrarily closely because the reliable resolution of the sharper and sharper 
quasiparticle peak would require a better and better resolution. This in turn 
cannot be achieved with a reasonable amount of computing resources because 
Eq.~\ref{eq:fsites_broadening} must be respected. For this reason,
we have to resort to an extrapolation to determine $U_{\text{c2}}$
quantitatively.

The behavior of $Z(U)$ close to $U_{\text{c2}}$ was already investigated
previously analytically by the projected self-consistent method (PSCM)
\cite{Moeller-1995} and by the linearized DMFT.\cite{Bulla-2000} 
Both methods assume the  separation of energy scales and
predict a linear behavior of $Z$ close to $U_{\text{c2}}$.
Since our all-numerical spectral densities strikingly confirm the separation of
the energy  scales \cite{Karski-2005} we determine  $U_{\text{c2}}$ by fitting
\begin{equation}
  f_{Z}(U) = a_1(U-U_{\text{c2}}) + a_2(U-U_{\text{c2}})^2
\end{equation}
to the data points for $U\in[2,\,2.8]D$. We obtain a critical value of
$U_{\text{c2}}=(3.07\pm0.1)D$ which agrees excellently within the error range
with most of the previous results,\cite{Moeller-1995, Bulla-1999, Bulla-2000,
  Bulla-2001, Garcia-2004,blume05a,Bluemer-2005b} including the linearized
DMFT\cite{Bulla-2000} and the PSCM.\cite{Moeller-1995} For comparison, we
include an overview of these critical values in Tab.~\ref{tab:u_critical}.

\begin{table}[ht]
  \centering
  \begin{tabular}{c|c|l}
    $U_{\text{c1}}$/D & $U_{\text{c2}}$/D & method\\
    \hline
    \hline
    $2.38  \pm 0.02$  & $3.07\pm 0.1$  & \mbox{D-DMRG} (CV, direct inversion)\\
    \hline
    $2.225 \pm 0.025$ & $3.05 \pm 0.05$ & \mbox{D-DMRG} 
    (CV, variational)\cite{Nishimoto-2004a,nishi06}\\
    $2.39  \pm 0.02$  & $3.0 \pm 0.2$  & \mbox{D-DMRG} (Lanczos)\cite{Garcia-2004}\\
    $2.39$            &                & PT-QMC\cite{Bluemer-2005b}\\
    $2.39$            & $2.94$         & NRG\cite{Bulla-1999, Bulla-2001}\\
                      & $3$            & linearized DMFT\cite{Bulla-2000}\\
                      & $2.92\pm 0.05$ & PSCM\cite{Moeller-1995}\\
    \hline
  \end{tabular}
  \caption{\label{tab:u_critical}
    Comparison of the values for
    $U_{\text{c1}}$ and $U_{\text{c2}}$ found by various approaches.}
\end{table}

\subsubsection{Momentum-Resolved Spectral Densities}
\label{sssec:met:momres}

\begin{figure}[ht]
  \includegraphics[width=\columnwidth]{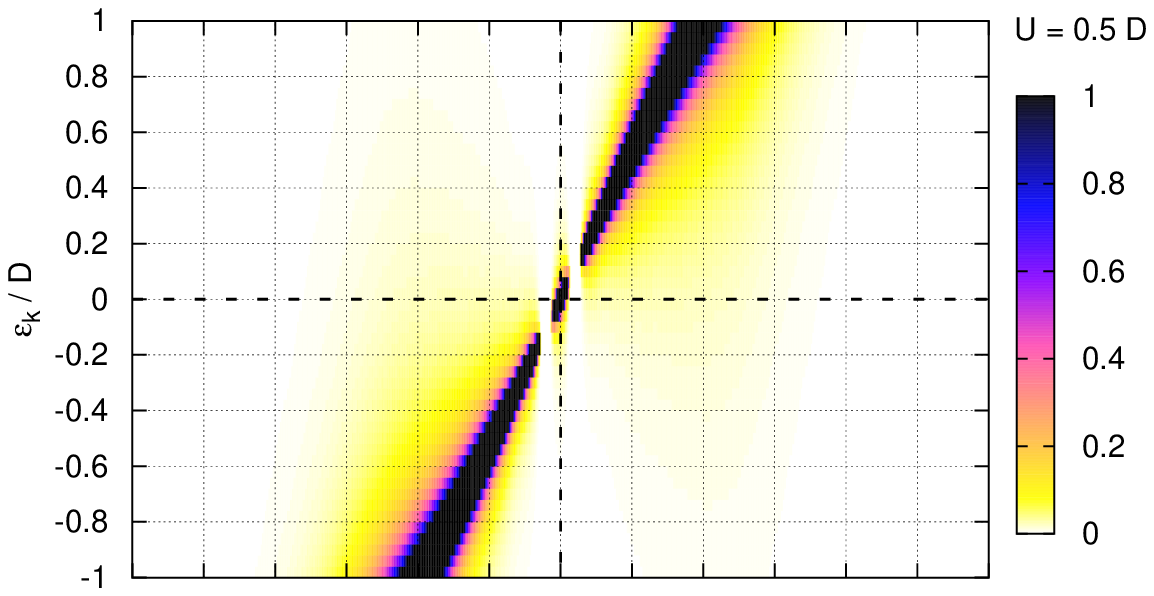}
  \includegraphics[width=\columnwidth]{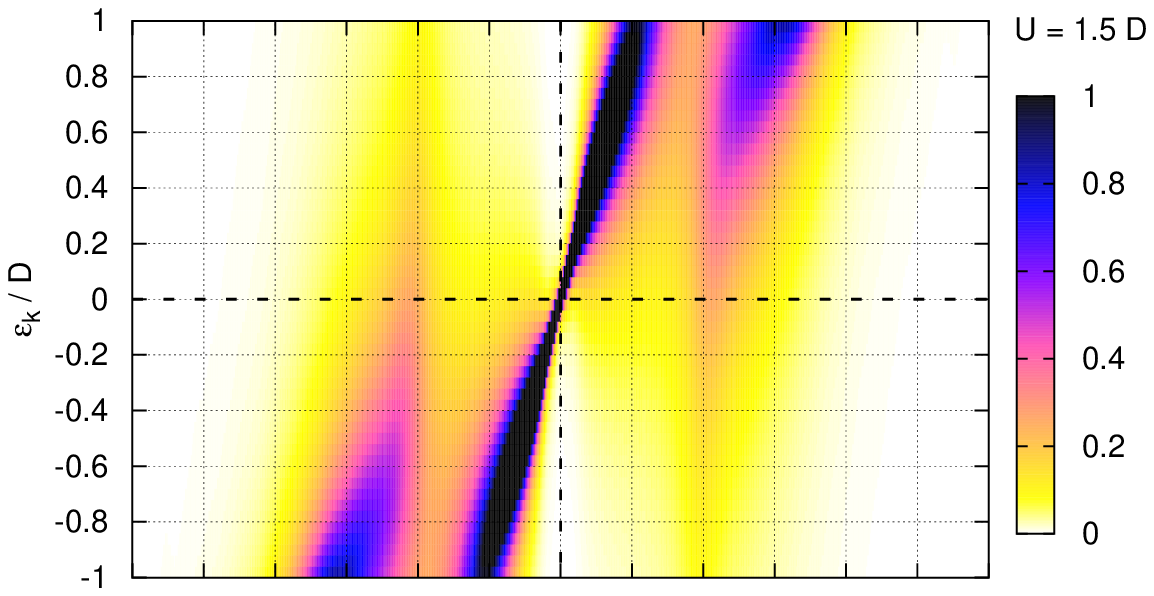}
  \includegraphics[width=\columnwidth]{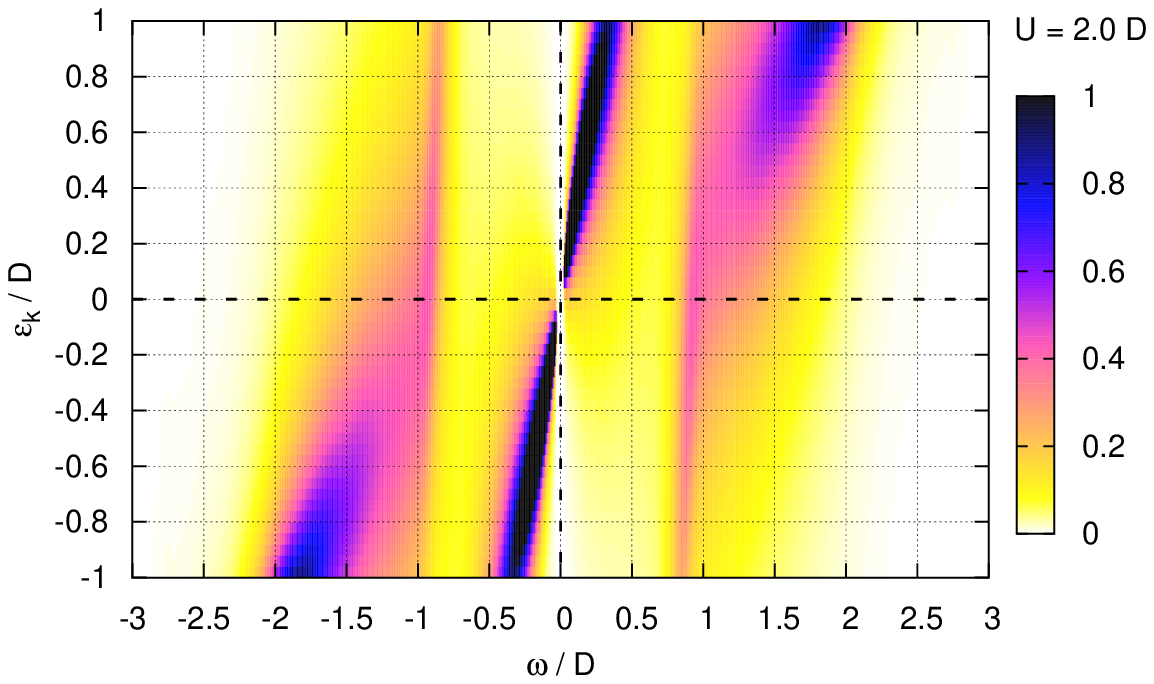}
  \caption{\label{fig:met:1}
    (Color online)
    Spectral densities $\rho(\omega, \epsilon_k)$ in the metallic regime for
    $U=0.5D$ (top), $U=D$ (middle), and $U=2D$ (bottom).}
\end{figure}

\begin{figure}[ht]
  \includegraphics[width=\columnwidth]{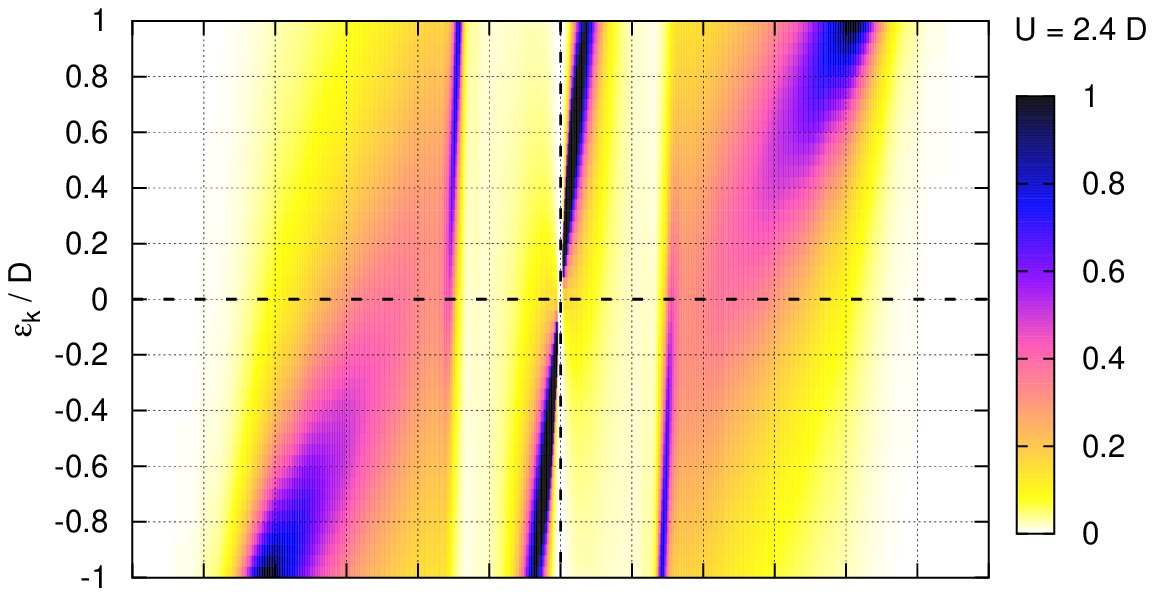}
  \includegraphics[width=\columnwidth]{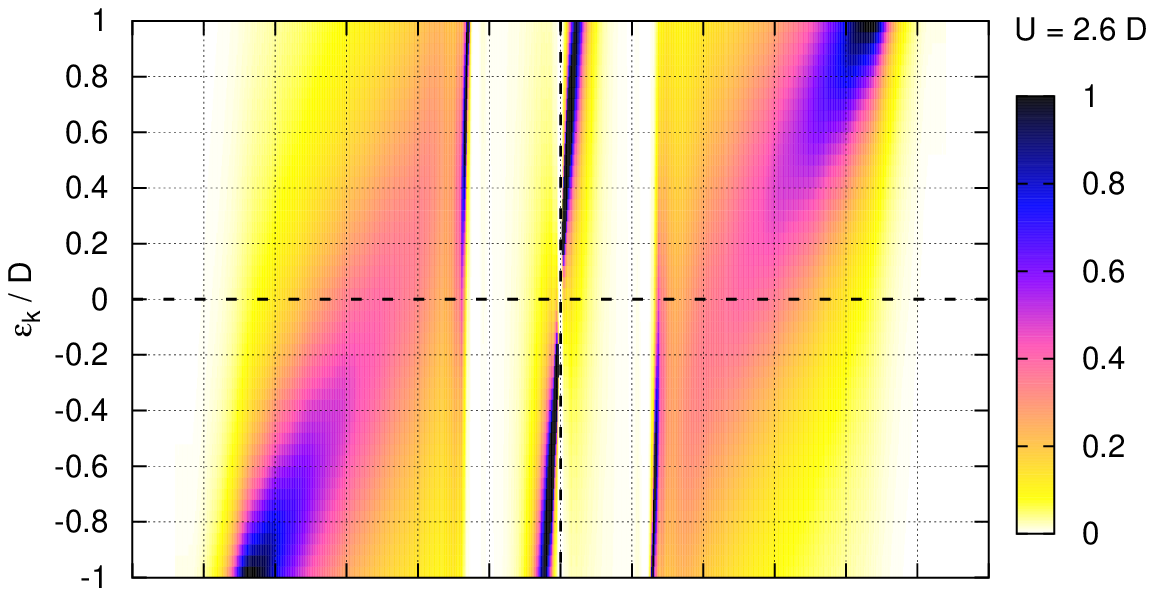}
  \includegraphics[width=\columnwidth]{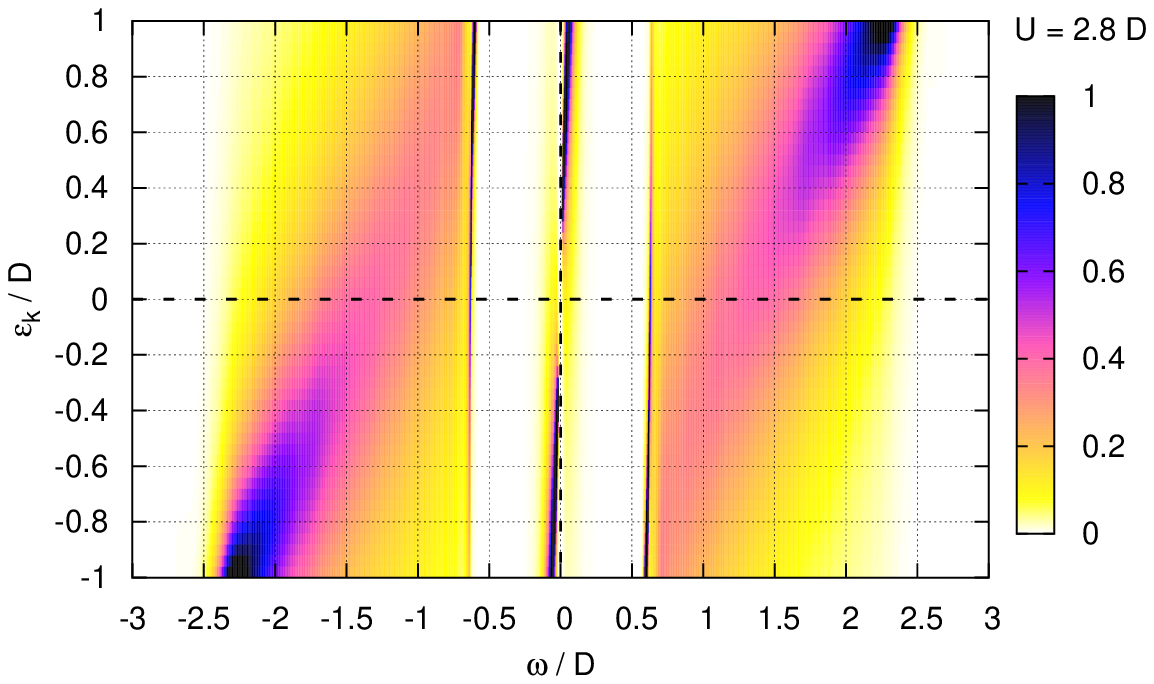}
  \caption{\label{fig:met:2}
    (Color online)
    Spectral densities $\rho(\omega, \epsilon_k)$ in the metallic regime for
    $U=2.4D$ (top), $U=2.6D$ (middle), and $U=2.8D$ (bottom).}
\end{figure}

For the metallic phase, we present spectral densities 
$\rho(\omega,\epsilon_k)$ as defined
in Eqs.\ (\ref{eq:gkomega},\ref{eq:rhokomega})
which are resolved according to their bare single-particle energy
$\epsilon_k$. In the context of the DMFT, this procedure provides 
information on the momentum dependence of the spectral properties.
Figs.\ \ref{fig:met:1} and  \ref{fig:met:2} display how the
momentum-resolved spectral densities evolve on approaching the 
metal-to-insulator transition.


Three main features can be discerned. First, the dominant diagonal
line of highest intensity with $\omega_\text{high} \propto \epsilon_k$
catches our attention in all six panels. This is where the 
Fermi liquid picture manifests itself.\cite{nozie97}
Without interaction a quasiparticle can be excited at
 $\omega_\text{high} = \epsilon_k$. In the presence of interaction
there are still quasiparticles which behave like fermions above the
Fermi level or holes below it. But these excitations are dressed by
clouds of particle-hole pairs. This dressing has two effects. One
is that the bare fermionic operator has only an overlap
reduced by $Z$ with the creation of the quasiparticle.  
The other effect is that the dressed quasiparticles are heavier
than their bare counterparts. This means that their dispersion is
reduced. For a momentum-independent self-energy the reduction 
of the dispersion is given again by the factor $Z\le 1$.
This can be seen in the impressive steepening of the dominant diagonal
line  given by $\omega_\text{high} = Z \epsilon_k$ upon increasing
interaction. For interaction values close to the instability of
the metallic solution, see lowest panel in Fig.\ \ref{fig:met:2},
the diagonal line has become so steep that it can almost be taken as
a vertical one. This implies that the quasiparticle has become almost
local and non-dispersive. This is the mechanism leading to the
so-called heavy fermions.

We point out that the white stripes at $\omega=0$ in most of the panels
in Figs.\ \ref{fig:met:1} and  \ref{fig:met:2} and the white stripes
at small $\omega$ in the first panel of  Fig.\ \ref{fig:met:1}
result from the numerical violations of causality as discussed before,
see Sects.\ \ref{sssec:ins:selfenergy} and \ref{sssec:met:selfenergy}.
For values of $\omega$ where this problem occurs the spectral densities
are set to zero.

The second main feature in the momentum-resolved spectral densities
is the development of the precursors of the lower and the upper Hubbard
band. At $U=0.5D$ no such precursors are visible, except for some 
broadening of the diagonal line away from the Fermi level. But in the
panel for $U=1.5D$ the precursors are already clearly visible. They are
qualitatively very similar to the broad densities found in the
insulating spectral densities in Figs.\ \ref{fig:iso:1} and \ref{fig:iso:2}.
This similarity becomes more and more pronounced as $U$ approaches
$U_\text{c2}$ where the metal changes continuously to the insulator.

The third main feature is the occurrence of a vertical line of high
density at the inner edges of the precursors of the Hubbard bands.
It is barely visible for $U=2.0D$, see last panel of Fig.\ \ref{fig:met:1}.
But it is clearly discernible for the larger values of the interaction
displayed in Fig.\ \ref{fig:met:2}. This vertical line represents the
momentum-resolved feature of the sharp side peaks at the inner
band edges discussed in Sec.\ \ref{sssec:met:sisp}. We re-emphasize 
that there is no discernible momentum dependence indicating a local, 
non-dispersive, immobile excitation.

We draw the reader's attention to the fact that the vertical lines
display an interesting dependence of their weight on the momentum.
The high density is found for $\omega >0$ for $\epsilon_k <0$ and
vice versa ($\omega <0$ for $\epsilon_k >0$).  This implies that the
immobile composite entity is excited upon adding a bare fermion ($\omega>0$),
but for momenta corresponding to occupied states ($\epsilon_k <0$) and 
vice versa. We deduce that the addition of the bare fermion
must induce the creation of a quasiparticle hole, cf.\ Sec.\ 
\ref{sssec:met:sisp}. Further work to elucidate this puzzling behavior
is called for.


\section{Summary}
\label{sec:summary}

The present work constitutes the comprehensive analysis of the
transition from a metal to an insulator driven by strong repulsive
interaction. The framework of this investigation is the dynamic
mean-field theory (DMFT). We evaluate the necessary self-consistency
equations at zero temperature with a numerical impurity solver
based on dynamic density-matrix renormalization (D-DMRG). The
D-DMRG computes a frequency-dependent correction vector which ensures
a high reliability of the results at all energies,  not only
close to the Fermi level. Thereby, we are able to provide
spectral densities in the metallic and in the
insulating phase with unprecedented resolution. Such densities
are relevant to photoelectron spectroscopy of transition metal
compounds and other narrow-band systems.

The calculations are done for a semi-elliptic non-interacting
density of states (DOS).
All the important technical details of the intricate combination
of D-DMRG and DMFT are given in detail so that our findings
may be reproduced independently.

The local, momentum-integrated, spectral density in the insulating phase is 
governed by the lower and the upper Hubbard band. Both are featureless
and display square-root singularities at the inner band edges. The
insulating gap $\Delta$ vanishes continuously for 
$U\to U_\text{c1}=(2.38 \pm 0.02)D$. A slight curvature indicates that
$\Delta \propto (U- U_\text{c1})^\zeta$ with $\zeta=0.72\pm 0.05$.

For $U<U_\text{c2}=(3.07\pm0.1)D$ the insulating solution is only metastable.
The phase with lower ground-state energy is the metallic one.\cite{Karski-2005}
The corresponding spectral densities show three important features for 
significant interaction values:
(i) the central quasiparticle peak,
 (ii) the precursors of the lower and the upper Hubbard band, and
(iii) the sharp side peaks at the inner band edges of the Hubbard bands.

The DOS of the central peak at the Fermi level is pinned 
to its non-interacting value due to Luttinger's theorem. The central
peak is built from heavy fermions or holes which disperse only weakly due to 
their strong dressing. They are the characteristic signature of a Fermi liquid.

The Hubbard bands signal that the metal evolves towards an insulator
on increasing interaction. Our findings, both the earlier ones 
\cite{Karski-2005} and the present ones, clearly show that the metallic 
spectral densities evolve continuously to the insulating one upon
$U\to U_\text{c2}$. This continuity holds in the sense of an integral norm,
i.e., considering the distribution of spectral weight,
not in the sense of an absolute value norm which jumps.
For instance, the DOS at the Fermi level is $2/(\pi D)$ in the metallic
regime and zero in the insulating one. But the corresponding \emph{weight}
vanishes linearly for  $U\to U_\text{c2}$. The pseudogap develops
simultaneously to the insulating gap.
These findings numerically confirm the assumption of the separation of energy 
scales.\cite{Moeller-1995,kotli99}

The sharp side peak at the inner edges of the Hubbard bands in the metallic
regime are the signature of a complex composite excitation which is
not yet understood. We found that its weight scales with the 
quasiparticle weight which supports our hypothesis that the side peaks
stem from an entity made from a quasiparticle and some collective,
immobile mode. The analysis of the momentum-resolved spectral densities
shows in addition that the involved quasiparticle is a hole if
a bare electron is added. Conversely, it is a particle if a bare hole is
created, i.e., an electron is taken out.

Ongoing research is devoted to the nature of the composite excitation
which generates the sharp side peaks. To this end, the response
functions, which correspond to various collective modes, 
need to be investigated.

Other interesting issues concern the qualitative and quantitative
dependence of the important features found on the non-interacting 
density of states. The changes introduced by doping certainly
will be of particular interest. A first study, based on Lanczos-DMRG,
of the effects of doping has been published very recently.\cite{garci07}


\begin{acknowledgments}
  We thank R.~Bulla for the kind provision of the NRG data and F.~B.~Anders, 
  P.~Grete, S.~Kehrein, and A.~Lichtenstein for helpful discussions. 
  The DFG is thanked for financial support in SFB 608 during the beginning of
  this project.
\end{acknowledgments}


\begin{appendix}


\section{Calculation of the Continued Fraction Coefficients}
\label{app:cf}

Once a propagator $g(\omega)$ is known on a dense mesh of frequencies 
(about $10^{5}$ points), one is able to determine its
continued fraction representation to an arbitrary depth
\begin{equation}
\label{eq:cfc_rep_green}
  g(\omega) = \cfrac{{b}_0^2}{\omega - a_0 -
    \cfrac{b_1^2}{\omega - a_1 -
      \cfrac{b_2^2}{\omega - a_2 - \cdots}}}
\end{equation}
very precisely. Numerically, only integrations are involved.

For the iteration, we define recursively the resolvents
\begin{subequations}
\begin{align}
  p_0(\omega) & := g(\omega)\ ,\\
  \label{eq:g_n}
  p_{n+1}(\omega) & = \omega-a_n-\frac{b_n^2}{p_n(\omega)}.
\end{align}
\end{subequations}
Obviously, the continued fraction representation of $p_n(\omega)$ 
starts with $a_n$ and $b_n$. It is shifted by $n$ relative to the
continued fraction of $g$. An expansion in $1/\omega$ reads
\begin{equation}
\label{eq:cfc-expansion}
p_n(\omega) = \frac{b_n^2}{\omega}\left(1+\frac{a_n}{\omega} 
+\mathcal{O}(\omega^{-2}) \right).
\end{equation}
Comparing this expression with the  expansion in $1/\omega$ 
of the Hilbert representation of $p_n(\omega)$
\cite{Pettifor-1985,Viswanath-1994} leads to 
\begin{subequations}
\label{eq:cfc}
\begin{align}
  \label{eq:b_n}
  b_n^2 & = -\frac{1}{\pi}\int\limits_{-\infty}^{\infty}
  \mathrm{d}\omega\,\mathfrak{Im}\,p_n(\omega)\ ,\\
  \label{eq:a_n}
  a_n & = -\frac{1}{\pi b_n^2}\int\limits_{-\infty}^{\infty}
  \mathrm{d}\omega\,\omega\,\mathfrak{Im}\,p_n(\omega).
\end{align}
\end{subequations}
The iteration of the above equations for $n= 0,1,2, \ldots$
yields the wanted coefficients. The iterations are stopped once
all the coefficients for the numerically tractable
finite chain of $L$ fermionic sites are determined, i.e., for
$n=L-3$.

In the case of a particle-hole symmetric model, the spectral density is
symmetric. Consequently, all odd moments and thus all the coefficients
$a_n$ ($n=0, 1, \ldots$) vanish in the above recursion scheme.
So for the model under study we need only to consider the equations
for the $b_n$.

The asymptotic behavior  of the continued fraction coefficients for 
$n\to\infty$ is determined by the band width of the spectral 
density $\rho(\omega)$ and by the singularities at the band edges.
\cite{Pettifor-1985,Viswanath-1994} We exploit the relation 
for the band width $2 B$, 
if the support of the spectral density is compact, which reads
\begin{equation}
\label{eq:gapless}
  4 |b_{\infty}| = 2 B
\end{equation}
where $b_{\infty}:=\lim_{i\to\infty}b_i$. 
For a gap $\Delta$ in the middle of the spectral density, 
i.e., for the insulator, the $b_i$ are oscillating for large $i$ like
\begin{equation}
\label{eq:gapful}
  b_i \to b_{\infty} \pm (-1)^i\Delta/2
\end{equation}
where $b_{\infty}$ is again given by a quarter of the band width which is
the distance between the largest and the smallest band edge including the gap.

In a strict sense we expect that the support of the spectral density
of an interacting system is infinitely large. 
But it turns out that significant spectral weight is found
only in a finite interval. The contributions outside this interval
are negligible which has been observed also in other calculations. 
\cite{Eastwood-2003}

We use the relations (\ref{eq:gapless},\ref{eq:gapful}) to determine the
interval in which we have to compute raw data $\{g_i\}$. 
For this purpose we compute $b_{\infty}$ in each iteration of the 
self-consistency cycle and observe its development on iteration. 
If the value of $b_{\infty}$ does not change significantly for several 
iteration steps, we fix the border $B$ of the interval in which we perform the 
calculation of the raw data.  This procedure allows us to
avoid the time-consuming calculation of unnecessary data points.

In the presence of an insulating gap $\Delta$ at the Fermi energy the 
above recursion scheme \eqref{eq:cfc} to determine the
continued fraction coefficients has to be modified for practical reasons. 
Due to the antisymmetry
of its real part the propagator in the particle-hole symmetric case
vanishes at $\omega=0$ completely $p_0(\omega=0)=0$ with no imaginary part. 
So there is a pole in $p_1(\omega)$ at $\omega=0$ according to \eqref{eq:g_n}
implying a finite contribution from a $\delta$-distribution  at $\omega=0$ in
$\mathfrak{Im}\,p_1$. We account for it separately
\begin{align}
  \begin{split}
    b_1^2 &=
    -\frac{1}{\pi} \int\limits_{-\infty}^{\infty}
    \mathrm{d}\omega\,\mathfrak{Im}\,p_1(\omega)\\
    &= -\frac{1}{\pi}\int\limits_{\mathbb{R}\setminus[-\Delta,\Delta]}
    \mathrm{d}\omega\,\mathfrak{Im}\,p_1(\omega)
    +\frac{b_0^2}{|p'_0(0)|}.
  \end{split}
  \label{eq:recur-pole}
\end{align}
The recursion relation \eqref{eq:g_n} implies that the same scenario
occurs for every odd $n$. So we use \eqref{eq:recur-pole} for any odd $n$ once
the subscript $_1$ is replaced by $_n$ and the subscript $_0$ by $_{n-1}$.


\section{Removal of Deconvolution Artifacts}
\label{app:gap}

\subsection{Outer Band Edges}

\begin{figure}[ht]
  \centering\includegraphics[width=\columnwidth]{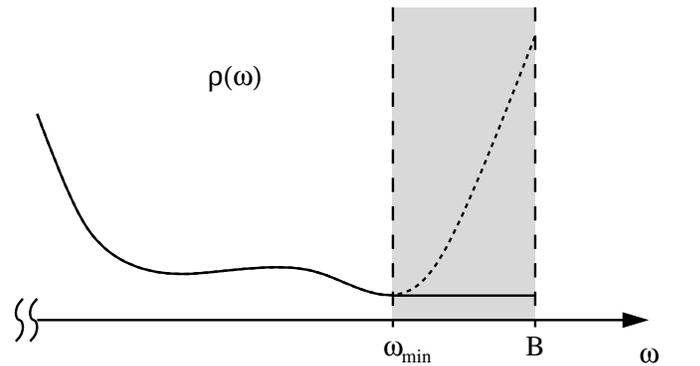}
  \caption{\label{fig:edge-artifact}
    Lines: Generic deconvolved spectral density close to the band
    edge at $B$. The dashed line marks the artifact which is removed
    by setting $\rho(\omega>\omega_\text{min}) = \rho(\omega_\text{min})$.
    The left band edge is treated accordingly.}
\end{figure}

The raw data $\{g_i\}$ from D-DMRG covers only a certain finite interval. 
Hence the underlying unbroadened spectral density $\rho(\omega)$
can also be determined only on a finite interval $[-B,B]$ ($B\approx 3D$) 
which is very close to the interval in which the raw data is available.
The deconvolution leads to an abrupt increase of the deconvolved
$\rho(\omega)$ at the assumed  band edges at $B$,
see Fig.\ \ref{fig:edge-artifact}. This increase is definitely an
artifact. In general, we expect that the spectral density $\rho(\omega)$ 
continues to decrease for $|\omega| \to \infty$. This decrease is a gradual
one, i.e., there is no band edge in the strict sense for the correlated 
system. But all results obtained so far by us and others \cite{Eastwood-2003} 
point into the direction that the contributions beyond a certain interval
can be safely neglected.

The artificial increase is removed by replacing $\rho(\omega)$ for
frequencies beyond $\omega_\text{min}$ where the first minimum occurs
by the minimal value $\rho(\omega_\text{min})$. This is indicated by
the dashed line in Fig.\ \ref{fig:edge-artifact} which is replaced by
the solid curve.

\subsection{Inner Band Edges of the Insulator}
\label{sapp:inner}

\begin{figure}[ht]
  \centering\includegraphics[width=\columnwidth]{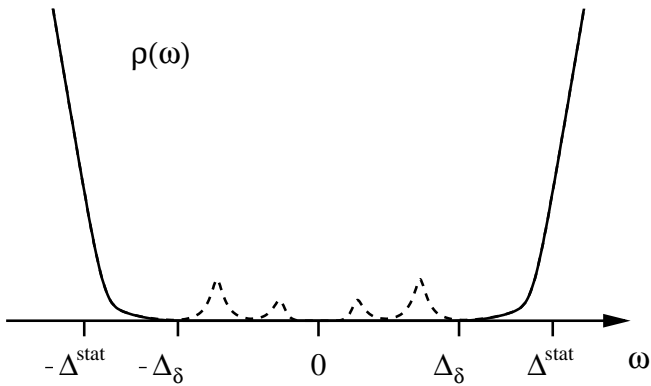}
  \caption{\label{fig:gap-artifact}
    Lines: Generic deconvolved spectral density in the region of the
    gap $\Delta$ of the insulating phase. The dashed line marks
    the artifacts which are cut out by setting $\rho(\omega)$ to zero for
    $\omega\in[-\Delta_\delta,\Delta_\delta]$.}
\end{figure}

The situation close to inner band edges stemming from a gap
is similar to the one close to outer band edges.
The LB deconvolution of the D-DMRG raw data in the region, where
the insulating solution displays a gap, leads to small artifacts
as shown in Fig.\ \ref{fig:gap-artifact} by the dashed lines.

The first remedy, which one may consider, is to determine the gap 
$\Delta^\text{stat}$ of the SIAM by conventional 
static DMRG and to use this value as 
the gap of the lattice problem in DMFT. But we found out that even an 
extrapolation of this static gap $\Delta^\text{stat}$ to the thermodynamic 
limit $L\to\infty$, as proposed in Ref.~\onlinecite{Nishimoto-2004a}, 
cannot be used in our iterations. No stable self-consistent solution 
in the sense of the criteria given in Sec.\ \ref{sssec:sc} could be obtained.

The successful treatment of the artifacts consisted in removing
only the obviously artificial features which are depicted
by dashed lines in Fig.\ \ref{fig:gap-artifact}. The solid lines are
kept because they represent real spectral weight. This is done by
starting from the value of the static gap $\Delta^{\text{stat}}$
adjusted by a correction parameter $\delta>0$
\begin{equation}
  \rho(\omega) \equiv 0 \quad
  \forall\omega\in[-\Delta^{\text{stat}} +\delta,\,
  \Delta^{\text{stat}} - \delta].
\end{equation}
The value $\delta$ is  chosen carefully, so that on the one hand all 
deconvolution artifacts are removed and on the other hand the proper 
spectral weight of the Hubbard bands stays untouched. 
In order to determine the appropriate value of $\delta$ it is
helpful to resolve the inner Hubbard band-edges in detail, 
which can be achieved by small values of $\eta_i$ in this region. 
Note, that the adjusted interval $[-\Delta^{\text{stat}} +\delta,\,
  \Delta^{\text{stat}} - \delta]$ can be set smaller than the interval 
$[-\Delta, \Delta]$ defined by the proper gap $\Delta$ as
long as all deconvolution artifacts are removed.
The artifacts would spoil the DMFT self-consistency iteration because
the continued fraction coefficients could not be evaluated properly.

It turns out that $\delta$ does not need to be changed in each iteration of the
DMFT self-consistency cycle. It is sufficient to set it once for each
value of $U$.
\end{appendix}

\vfill



\end{document}